\documentclass[fleqn,10pt]{olplainarticle}

\title{A High-Order Compact Finite Volume Method for Unstructured Grids: Scheme Space Formulation and One-Dimensional Implementations}

\author[1]{Ling Wen}
\author[2]{Yan-Tao Yang}
\author[3]{Qing-Dong Cai}
\affil[1,2,3]{State Key Laboratory for Turbulence and Complex Systems, Department of Mechanics, 	School of Mechanics and Engineering Science,	Peking University, Beijing, 100871, China}


\begin{abstract}
This paper presents a novel and straightforward compact reconstruction procedure for the high-order finite volume method on unstructured grids.  In this procedure, we constructed a linear approximation relationship between the mean values and the function values, as well as the derivative values. Compared with  the classical compact schemes, which employ a Taylor expansion method to determine the coefficients, our approach adopts an equivalent and more generalized method to achieve this goal.  Via this method, the problem of constructing a high-order compact scheme is transformed into solving the null space of underdetermined homogeneous linear systems.
This null space constitutes the complete set of schemes that meet the specified accuracy under a given stencil, and is termed the ‘scheme space’
Schemes within the scheme space possess the same accuracy level yet exhibit distinct dispersion and dissipation characteristics.
Through  Fourier analysis, we can get the dissipation and dispersion properties of all schemes in the scheme space. 
This facilitates the control of scheme dispersion and dissipation without altering the stencil compactness
Combined with the WENO (Weighted Essentially Non-Oscillatory) concept, multi-stencil schemes are employed to construct the nonlinear weighted compact finite volume scheme (WCFV).
The WCFV is capable of eliminating unphysical oscillations at discontinuities, thereby enabling the capture of strong discontinuities.
One-dimensional schemes are discussed in detail, and numerical results demonstrate that the proposed method exhibits high-order accuracy, robustness, and shock-capturing capability.
\end{abstract}

\begin{document}

\flushbottom
\maketitle
\thispagestyle{empty}

\section{Introduction}

Compact methods are a class of high-order accurate methods that achieve high-order accuracy while minimizing the demand for stencil points or elements. These methods typically offer higher accuracy and resolution on the same computational stencil, along with smaller dissipation errors. Compared with low-order methods, high-order compact methods yield more accurate solutions at equivalent computational cost and exhibit higher efficiency in simulating flows requiring high accuracy \cite{ZINGG2000683}.  Compared with traditional high-order methods, compact methods reduce the number of stencils—translating to less information exchange between different CPUs or GPUs during parallel computation, and thus enhancing computational efficiency.  The advantages of high-order compact methods have attracted attention and research.  Compact methods were first applied in finite difference methods, where the core idea is to express the finite difference approximation of a function’s derivative as a linear combination of given function values at a set of nodes. The work that popularized compact methods is attributed to Lele \cite{lele1992compact}. Lele derived a series of central compact schemes (CCS) for the one-dimensional finite difference method. Fourier analysis shows that these compact schemes exhibit spectral-like resolution for short waves. CCS are straightforward to construct and can achieve high accuracy with a relatively small number of grid points. However, they exhibit limitations when applied to numerical simulations of various physical problems. First, these schemes exhibit insufficient numerical dissipation, leading to inherent numerical instability in long-term simulations. Additionally, when pursuing higher accuracy, the number of grid points remains large, complicating the construction of schemes for  points near boundaries. Furthermore, when CCS are applied to problems involving discontinuities, unphysical oscillations emerge.

Building on Lele’s work, compact methods have been further studied and extended to various physical problems. Deng developed a class of single-parameter linear dissipative compact schemes (DCS) \cite{deng1996class}, which introduce dissipation terms into CCS to enhance the stability of compact schemes. By adjusting the parameter value, the numerical dissipation of DCS can be controlled. To mitigate unphysical high-frequency oscillations near discontinuities, various upwind compact schemes have been developed to solve nonlinear problems involving discontinuities \cite{yamamoto1993higher,zhong1998high,ZHOU20071306,CHAMARTHI2021110067}. Compared with CCS, upwind compact schemes introduce additional numerical dissipation. This ensures stable solutions when solving problems with discontinuities. Additionally, to further improve scheme accuracy without increasing the number of stencil points, Fu et al. proposed the super compact finite difference method (SCFDM) \cite{ma1996super,dexun2001analysis}. In SCFDM, compact schemes are constructed using function values and the values of various orders of derivatives at stencil points. To further suppress unphysical oscillations at discontinuities, the concepts of ENO (Essentially Non-Oscillatory) \cite{shu1988efficient, shu1989efficient2, abgrall1994essentially} and WENO \cite{friedrich1998weighted} have been incorporated into compact schemes. Deng et al. proposed nonlinear weighted compact finite difference schemes (WCNS) \cite{deng2000developing, deng2010extending}, which have been successfully applied to a broad range of flow problems. Liu et al. \cite{liu2013new,liu2015new} proposed a new class of central compact schemes, which were subsequently extended to hybrid weighted nonlinear schemes. By nonlinearly weighting multiple linear compact schemes, oscillations near discontinuities can be largely eliminated.  This enables the application of high-order compact schemes to problems with discontinuities.

\begin{figure}[htbp]
	\centering
	\includegraphics[height=4.5cm]{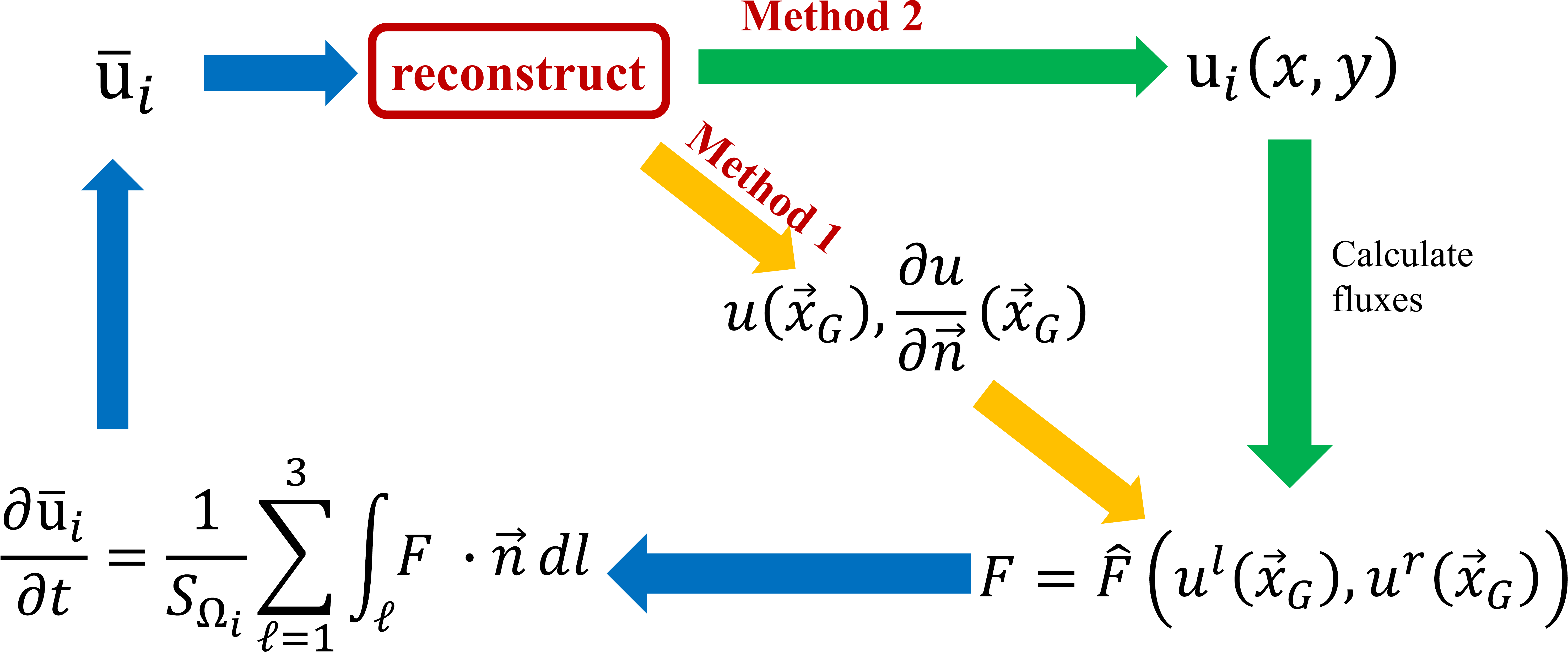}
	\caption{A general flow chart of finite volume method.}
	\label{fig_finite_volume_flowchart}
\end{figure}

The concept of compactness was soon extended to finite volume methods. In classical finite volume methods—such as the k-exact \cite{barth1990higher,ollivier2002high} and WENO \cite{friedrich1998weighted, hu1999weighted, dumbser2007arbitrary, dumbser2007quadrature} methods, polynomials for target elements are directly reconstructed using the averages of multiple  elements. This approach necessitates the use of large stencils to achieve high accuracy. Constructing compact finite volume schemes addresses the issue of large stencils. Fig. \ref{fig_finite_volume_flowchart} presents the flowchart for solving general conservation laws via the finite volume method. The core of high-order finite volume methods lies in approximating fluxes with high-order accuracy using element mean values. Based on the different approaches to flux calculation, the strategies for constructing compact finite volume methods can be categorized into two types.

The first type is analogous to the approach used for constructing compact finite difference methods. Its core idea is to linearly approximate function values and certain orders of derivatives at specific points (typically Gaussian integration points or grid nodes) using element mean values, followed by deriving approximate expressions for fluxes. Gaitonde and Shang first introduced a high-order finite volume method based on compact difference schemes in \cite{gaitonde1997optimized}.  Subsequently, Lacor et al. \cite{lacor2004finite} developed a central compact finite volume scheme suitable for multidimensional non-uniform structured  grids.
Fosso et al. \cite{fosso2010curvilinear}  further extended compact finite volume schemes to curvilinear coordinate griss via coordinate transformation. This method builds on the idea of compact finite difference methods to construct  schemes and has successfully extended their application to structured grids.  However,  this method faces substantial challenges when applied to unstructured grids. One key challenge is the unpredictable topology of unstructured grids, which leads to variations in the number of elements and  points when constructing linear compact schemes. This unpredictability complicates the computation of function values and derivatives at target points. Another challenge stems from the diverse element shapes in unstructured grids, which makes deriving coefficients via the Taylor expansion method difficult.

To construct compact finite volume schemes on unstructured grids, an alternative approach can be employed. Specifically, this approach uses mean values of face-adjacent elements and introduces additional constraint relations to reconstruct polynomials within each element. Subsequently, fluxes can be computed using the reconstructed polynomials. To ensure compactness, all constraint relations are restricted to adjacent elements. Building on this concept, various compact finite volume schemes can be constructed on unstructured grids. Wang et al. \cite{wang2002spectral, wang2004spectral, liu2006spectral} introduced the spectral finite volume method (SV) , which partitions a large mesh into smaller sub-meshes and imposes constraints on these sub-meshes.  This approach enables the successful implementation of high-accuracy finite volume methods with a small stencil. Ren et al. \cite{wang2016compactone,wang2016compacttwo} developed a compact least-squares finite volume method for unstructured grids. In this method, the reconstructed polynomial and its derivatives within a control volume must conserve their averages across face-adjacent elements. Furthermore, they proposed a high-accuracy finite volume reconstruction approach using variational principles \cite{wang2017compact}, which can be used in arbitrary type of  unstructured grids. In our previous study \cite{WEN2025113461}, we adopted a similar approach to 
construct compact finite volume schemes. The key idea is to introduce control points on the common edges of adjacent elements and impose continuity constraints on certain orders of normal derivatives at these control points. Particularly, we proposed  a fourth-order compact reconstruction on triangular unstructured grids.

In this paper, we further extend the first approach for constructing compact finite volume schemes and present a unified framework for developing high-accuracy compact finite volume schemes from a null space perspective. The key aspect of this approach is to approximate function values and derivatives at specific points using mean values of a set of elements. In this procedure, we require that the number of variables used to construct compact schemes exceeds the number of variables required to achieve the desired accuracy. Diverging from classical methods, we transform the problem of constructing compact schemes into solving the null space of underdetermined homogeneous linear equations by adopting a procedure equivalent to the Taylor expansion method. Fortunately, well-established and straightforward mathematical methods exist for solving the latter problem. This enables the straightforward construction of a family of compact schemes. And both mathematical and numerical proofs confirm that following our procedure yields compact schemes with the desired accuracy. This enables the straightforward construction of a family of compact schemes, and both mathematical and numerical proofs confirm that following our procedure yields compact schemes with the desired accuracy. The null space consists of all schemes capable of achieving the designed accuracy for a given stencil, and each vector within the scheme space corresponds to a compact scheme. To better emphasize the characteristics of this space, we term it the ‘scheme space’. Schemes within the scheme space possess the same or higher accuracy but exhibit distinct dispersion and dissipation properties. Through the Fourier analysis, we can obtain the dissipation space and dispersion space corresponding to the scheme space. This allows control over scheme dissipation and dispersion by selecting different compact schemes within the scheme space. Building on these linear schemes, we integrate the WENO concept to propose nonlinear weighted compact finite volume schemes (WCFVM). The core procedure involves nonlinearly combining multiple compact schemes to suppress unphysical oscillations at discontinuities and capture sharp interfaces. In this paper, we primarily focus on the properties of the null space-based method for constructing compact schemes and discuss one-dimensional schemes built using this method in detail. Nevertheless, we emphasize that our method can be extended to 2D or 3D unstructured grids. The procedure for constructing compact schemes on unstructured grids is similar, and the problem is ultimately transformed into solving the null space of underdetermined homogeneous linear equations.  To illustrate this claim, we construct a simple fourth-order compact scheme for 2D triangular unstructured grids in Section 4 and verify its numerical accuracy in Section 5.

This paper is organized as follows. Section 2 details the basic concepts, formulations, Fourier analysis, and WCFV scheme development of the proposed high-order compact finite volume method. Section 3 introduces specific schemes, including 1D function/derivative approximation schemes, hybrid schemes, and a 2D fourth-order scheme for unstructured triangular grids. Section 4 verifies the method via numerical benchmarks like linear advection and shock tube problems. Finally, Section 5 summarizes key findings and outlines future research directions.

\section{High order compact finite volume method}
\subsection{Construction method}
This subsection presents the basic concepts and formulations of the proposed high-order compact schemes. For clarity, several notations are defined first.

The computational domain $\Omega$ is partitioned into $N$ non-overlapping elements, i.e.,
\begin{equation}
	\Omega=\bigcup_{i=1}^{N}\Omega_i,
\end{equation}
where $\Omega_i=\left[x_i,x_{i+1}\right]$ represents the $i$-th element. The mean value of the function $u\left(x\right)$ over each control volume $\Omega_i$ is defined as follows,
\begin{equation}
	{\bar{u}}_i=\frac{1}{h_i}\int_{\Omega_i} u\left(x\right)dx,
\end{equation}
where $h_i=x_{i+1}-x_i$ is the length of $\Omega_i$ and ${\bar{u}}_i$ denotes the mean value of $u\left(x\right)$ over $\Omega_i$.

The goal of a compact finite volume scheme is to use the given mean values of control elements to reconstruct function values or derivatives at specific points. In this context, the finite volume approximations to $u\left(x_i\right)$ (the function value) and $u^{\left(m\right)}\left(x_i\right)$ (the $m$-th order derivative) at these points are denoted by $u_i$ and $u_i^{\left(m\right)}$, respectively. The key concept of the proposed method is to express the values of certain orders of derivatives at stencil points in a linear equation using element mean values, i.e.,
\begin{equation}
	\bm{f}\cdot\bm{c}=R\left(O\left(h^{k+1}\right)\right).
	\label{eq2.3}
\end{equation}
Here $\bm{f}$ is a vector consisting of mean values of stencil elements and  the $m$-th order derivatives at stencil points and $\bm{c}$ is a vector of coefficients to be determined.  $h$ denotes the grid scale. $R$ represents the truncation error, and $k+1$ is the accuracy order of  Eq. (\ref{eq2.3}).

In Eq. (\ref{eq2.3}), a linear relationship with $(k+1)$-th order accuracy is established using mean values of the stencil elements $S={\Omega_1,\Omega_2,\ldots,\Omega_s}$ and the values of $m$-th order derivatives at the stencil points $Q^m=\left\{q_1^m, q_2^m, \ldots, q_{n_m}^m \right\}$, where $m=0,1,2,\ldots,M$. The specific forms of $\bm{c}$ and $\bm{f}$ are defined by
\begin{equation}
	\bm{c}=\left[c_1,c_2,c_3,\ldots,c_n\right]^T,
\end{equation}
and
\begin{equation}
	\bm{f}=\left[{\bar{u}}_1,{\bar{u}}_2,\ \ldots,{\bar{u}}_s,h^mu_{q_1}^{\left(m\right)},h^mu_{q_2}^{\left(m\right)},\ldots,h^mu_{q_{n_m}}^{\left(m\right)}\right],
	\label{eq2.5}
\end{equation}
respectively. Here, the superscript $"T"$ denotes the transpose of a vector. $s$ denotes the number of stencil elements,  and $n_m$ represents the number of stencil points at which the $m$-th order derivative is used. $\bm{f}$ is a row vector of size $1\times n$, where $n$ is the total number of variables used in the compact scheme, i.e.,
\begin{equation}
	n=s+\sum_{i=0}^{m}n_i.
	\label{eq2.6}
\end{equation}
Here, $\sum_{i=0}^{m}n_i$ represents the total number of stencil points across all derivative orders. In Eq. (\ref{eq2.5}), $h^mu_{q_{n_m}}^{\left(n\right)}$ is included as a variable instead of $u_{q_{n_m}}^{\left(n\right)}$ to ensure consistent dimensionality of variables, thereby making the coefficients $\bm{c}$ dimensionless. Additionally, it is important to note that $m$ can take multiple values. In such cases, a combination of derivatives of different orders at various control points is used to construct compact schemes.

In our proposed method, the procedures for constructing compact finite volume and compact finite difference schemes are similar. The key distinction lies primarily in the selection of the variable vector $\bm{f}$. Compact finite volume methods use mean values to reconstruct the values of other variables.  In this paper, the reconstructed variables are function values or derivatives at control points. Consequently, the variable vector $\bm{f}$ include mean values of specific elements.  In contrast, compact finite difference methods use function values at nodes to reconstruct other variables, such as derivatives or function values at points of interest. In this case, the variable vector must include function values at nodes. The core of constructing a compact scheme is to determine the coefficients $\bm{c}$ that achieve $k+1$-th order accuracy for the stencil $S$ and points $Q^m$. In classical compact methods \cite{deng1996class, ma1996super, dexun2001analysis, gaitonde1997optimized, lacor2004finite}, the coefficients $\bm{c}$ in Eq. (\ref{eq2.3}) are derived by matching Taylor series coefficients of different orders. In this paper, we adopt an alternative equivalent approach to derive the expression for the coefficients $\bm{c}$. 

Let $P^k$ denote the set of one-dimensional polynomials of degree $k$. The truncation error $R$ of Eq. (\ref{eq2.3}) is of order $\left(k+1\right)$, which implies that Eq. (\ref{eq2.3}) holds exactly for any polynomial of degree $k$ (i.e., $R=0$). This gives the following relation, i.e.,
\begin{equation}
	\bm{f}\cdot\bm{c}=0,\ \ \forall\ u\left(x\right)\in P^k.
	\label{eq2.7}
\end{equation}
The basis functions of the polynomial set $P^k$ are given by the vector $\bm{\varphi}\left(x\right)$, which is
\begin{equation}
	\bm{\varphi}\left(x\right)=\left[1,x,x^2,\ldots,x^k\right]^T.
	\label{eq2.8}
\end{equation}
All polynomials $u\left(x\right)\in P^k$ can be expressed as a linear combination of the basis functions in Eq. (\ref{eq2.8}). Therefore, Eq. (\ref{eq2.3}) achieves $(k+1)$-th order accuracy, which implies that $\bm{f}\cdot\bm{c}=0$ when $u(x)$ corresponds to any of the basis functions in $\bm{\varphi}\left(x\right)$. Substituting Eq. (\ref{eq2.8}) into Eq. (\ref{eq2.3}), we can obtain
\begin{equation}
	\mathbb{A}\bm{c}=0,
	\label{eq2.9}
\end{equation}
where $\mathbb{A}$ is a $\left(k+1\right)\times n$ matrix. The $i$-th column of the matrix $\mathbb{A}$ is defined as $\bm{a}_i,\ i=1,2,\ldots,n$, and the expression of $\bm{a}_i$ is
\begin{equation}
	\begin{aligned}
		& \bm{a}_i=\frac{1}{h_i}\int_{\Omega_i}\bm{\varphi}\left(x\right)dx,\ \ i=1,2,\ldots,s;\\
		& \bm{a}_{i+s}=h^m\bm{\varphi}^{\left(m\right)}\left(x_{q_{n_m}}\right),\ \ i=1,2,\ldots,n-s.
		\label{eq2.10}
   \end{aligned}
\end{equation}
For the $j$-th basis function $x^{j-1}$ of $\bm{\varphi}(x)$, we can obtain 
\begin{equation}
	\begin{aligned}
		& a_{i,j}=\frac{1}{h_i}\left(x_{i+1}^j-x_i^j\right),\ \ i=1,2,\ldots,s;\ j=1,2,\ldots,k+1;\\
		& a_{i+s,j}=\frac{\Gamma\left(j-1\right)}{\Gamma\left(j-m-1\right)}h^mx_{q_{n_m\ }}^{j-m-1},\ \ i=1,2,\ldots,n-s;\ j=1,2,\ldots,k+1.
		\label{eq2.10.new}
	\end{aligned}
\end{equation}
where  $a_{i,j}$ denotes the $(i,j)$-th element of matrix $\mathbb{A}$. $\Gamma\left(n\right)$ is the gamma function, defined such that $\Gamma\left(n\right)=0$ for n<0 and $\Gamma\left(n\right)=n!$ for $n\geq0$.

In deriving Eq. (\ref{eq2.9}), the polynomials of Taylor  expansion are chosen as the basis functions. In fact, any other set of convergent series expansion functions can be chosen as basis functions for Eq. (\ref{eq2.3}). For instance, the sine function family can be used as basis functions, corresponding to the Fourier series expansion of the function $u(x)$. For example, we can  use the sine function family as the basis function, which corresponds to the Fourier series expansion of the function. In this paper, only the case where polynomials serve as basis functions is considered.

To ensure the existence of non-trivial solutions, it is necessary that $k + 1 < n$. This implies that the highest degree of polynomial for which Eq. (\ref{eq2.3}) holds exactly is $n-2$, and the maximum achievable accuracy is $(n-1)$-th order. When $n>k+1$, i.e., the number of variables exceeds the number of basis functions, Eq. (\ref{eq2.9}) becomes an underdetermined $\left(k+1\right) \times n$ homogeneous linear system, and an infinite number of non-trivial solutions satisfying the conditions of Eq. (\ref{eq2.3}) exist. Our goal is to derive the general solution to Eq. (\ref{eq2.9}).

Through the aforementioned steps, the problem of constructing a compact scheme is transformed into deriving the general solution to an underdetermined homogeneous linear system. Fortunately, well-established and straightforward mathematical methods exist for solving this problem. Assume the rank of matrix $\mathbb{A}$ is $r$, where $r\le k+1<n$. Then, there are $n-r$ linearly independent solutions to Eq. (\ref{eq2.9}).  Let these $n-r$ linearly independent solutions be ${\bm{c}}_1, {\bm{c}}_2, \ldots, {\bm{c}}_{n-r}$. The general solution to Eq. (\ref{eq2.9}) can be expressed as a linear combination of these basis solutions, i.e.,
\begin{equation}
	\bm{c}=\eta_1\bm{c}_1+\eta_2\bm{c}_2+\ldots+\eta_{n-r}\bm{c}_{n-r}=\mathbb{C}\bm{\eta},
\label{eq2.11}
\end{equation}
where $\eta_i$ (for $i=1, 2, \ldots, n-r$) are arbitrary real coefficients, and $\bm{c}$ is the general solution. $\bm{\eta} = \left[\eta_1, \eta_2, \ldots, \eta_{n-r}\right]^T$ is a $\left(n-r\right)\times 1$ column vector, and $\mathbb{C}=\left[\bm{c}_1, \bm{c}_2, \ldots, \bm{c}_{n-r}\right]$ is an $n \times \left(n-r\right)$ matrix whose columns are the $n-r$ linearly independent solutions.

Eq. (\ref{eq2.11}) yields all schemes that can achieve $(k+1)$-th order accuracy using the variable vector $\bm{f}$. This space is referred to as the ‘scheme space’. The scheme space is a linear vector space. Any linear combination of basis solutions $\bm{c}_1,\ldots,\bm{c}_{n-r}$ belongs to  this space. It is important to distinguish between the linear combination (or linear space) mentioned here and a linear scheme. When the variable $\bm{\eta}$ depends on the variable vector  $\bm{f}$, the coefficients $\bm{c}$ also depend on $\bm{f}$. In this case, a nonlinear scheme is obtained. However, as long as $\bm{\eta}$ is not explicitly related to the basis solutions in $\mathbb{C}$, the coefficients $\bm{c}$ still lie within the space defined by Eq. (\ref{eq2.11}). This implies that, in smooth regions, the nonlinear scheme derived from Eq. (\ref{eq2.11}) still ensures $(k+1)$-th order accuracy.

Assume Eq. (\ref{eq2.3}) is used to solve for the $i$-th variable $f_i$ of $\bm{f}$. The corresponding coefficient $c_{i}$ is normalized, i.e.,
\begin{equation}
	\bm{c}^\prime=\frac{\bm{c}}{c_{i}}.
	\label{eq2.12}
\end{equation}
For simplicity and without ambiguity, $\bm{c}^\prime$ is still denoted as $\bm{c}$. In this paper, the relaxation iteration method is used to solve for $f_i$. Let $\alpha$ be the relaxation factor, and let the value at the $j$-th  iteration step be denoted by the superscript $j$. Then, we have
\begin{equation}
	\left(f_i\right)^{j+1}=\alpha\left(\bm{f}\cdot\bm{c}\right)^j+\left(f_i\right)^j.
	\label{eq2.13}
\end{equation}
The relaxation factor $\alpha$  typically ranges from 0 to 2. When $\alpha\in(1,2)$, an over-relaxation iteration is obtained. To ensure the convergence of the relaxation iteration,  appropriate coefficients vector $\bm{c}$ must be chosen. Specifically, the coefficient matrix formed by the coefficients vectors must be positive definite.  In this paper, the coefficients $c_i$ corresponding to the unknowns $c_i$ are chosen to be diagonally dominant, thereby ensuring iterative convergence. The relaxation iteration for reconstruction is terminated when the iteration residuals, defined as $\left|R^j\right| = \left|\left(\bm{f}\cdot\bm{c}\right)^j\right|$, are less than the specified tolerance, or when the maximum number of iterations is reached.

The following provides a concise proof of the following conclusion: compact schemes derived via the aforementioned reconstruction process achieve $k+1$-th order accuracy.

The $p$-th order Taylor expansion term of $u\left(x\right)$ at point $x_{i}$ is given by
\begin{equation}
	T_p\left(x\right)=u^{\left(p\right)}\left(x_{i}\right)\frac{\left(x - x_{i}\right)^p}{p!}, \ \ \  p=0,1,2,3,\ldots
	\label{eq2.14}
\end{equation}
Substituting Eq. (\ref{eq2.14}) into Eq. (\ref{eq2.3}), we can obtain the expression of the truncation error of the $p$-th order Taylor expansion term for the point $x_{i}$, given by $R_p=\bm{f}_p\cdot\bm{c}$. Here,  $\bm{f}_p$ denotes the variable vector $\bm{f}$ when $u(x)=T_p (x)$. From the preceding discussion, it follows that for $p<k+1$, the truncation error $R_p=0$, i.e.,
\begin{equation}
	R_p=0,\ \ p=0,1,2,\ldots,k.
\end{equation}
The total truncation error of Eq. (\ref{eq2.3}) is
\begin{equation}
	R=\sum_{p=0}^{\infty}R_p=\sum_{p=k+1}^{\infty}\bm{f}_p\cdot\bm{c} 
	=\sum_{p=k+1}^{\infty}\bm{f}_p\mathbb{C}\bm{\eta}.
	\label{eq2.16}
\end{equation}
In Eq. (\ref{eq2.16}), we can get $\bm{f}_p \sim O\left(h^p\right)$ when $u\left(x\right)=T_p\left(x\right)$, and $\bm{\eta}$ is bounded real vector, i.e. $\left|\eta_j\right|<\eta_0, j=1,2,\ldots,n-r$. This leads to the following relation,
\begin{equation}
	R=\sum_{p=k+1}^{\infty}\bm{f}_p\mathbb{C}\bm{\eta}\leq \eta_0 
	\left \| \mathbb{C} \right \|
	\sum_{p=k+1}^{\infty}\left \| \bm{f}_p \right \|  \sim O\left(h^{k+1}\right).
	\label{eq2.17}
\end{equation}
Here $\left \| (\cdot) \right \|$ denotes the norm of a matrix or  vector. The total truncation error $R$ satisfies $R \sim O(h^{k+1})$, which indicates that the aforementioned reconstruction process theoretically enables the derivation of a compact scheme with $k+1$-th order accuracy.

For each point $x_i,i=1,2,3,\ldots,N+1$,  the compact scheme in Eq. (\ref{eq2.3}) can be constructed. For non-uniform grids, the coefficients $\bm{c}$ of the schemes vary with the point. The truncation errors $R$ are  also different, but  satisfy $O\left(h^{k+1}\right)$. This property ensures that the schemes maintain their designed accuracy on non-uniform grids. For uniform grids, the coefficients $\bm{c}$ of the schemes are identical across all points. In the next section, we will analyze the dispersion and dissipation characteristics of different compact schemes for this case.

\subsection{Fourier analysis}
As discussed in the preceding subsection, schemes within the scheme space are equivalent in terms of accuracy, i.e., their truncation error $R$ satisfies $R\sim O(h^{k+1})$. However, these schemes exhibit distinct dispersion and dissipation characteristics. In this subsection, Fourier analysis is used to derive the Fourier mode errors of the compact finite volume scheme.  Specifically, the dispersion and dissipation characteristics corresponding to the scheme space are investigated for the one-dimensional scalar linear advection equation.

The scalar linear advection equation is
\begin{equation}
	\frac{\partial u}{\partial t}+\frac{\partial F}{\partial x}=0,
	\label{eq3.1}
\end{equation}
where $F=au$ denotes  the flux. $a$ is assumed to be a positive constant, i.e., $a>0$.  The integral form of Eq. (\ref{eq3.1}) over  control volume $\Omega_i=[x_i,x_{i+1}]$ is
\begin{equation}
	\frac{\partial{\bar{u}}_i}{\partial t}+\frac{1}{h_i}\left[F\left(x_{i+1}\right)-F\left(x_i\right)\right]=0,
	\label{eq3.2}
\end{equation}
where $F\left(x_i\right)=au\left(x_i\right)$ is the exact flux and $h_i=\left|x_{i+1}-x_i\right|$ is the length of $\Omega_i$.  The semi-discrete finite volume scheme approximates Eq. (\ref{eq3.2}) by introducing numerical fluxes $a u_{i+1}$ and $a u_i$, leading to
\begin{equation}
	\frac{\partial{\bar{u}}_i}{\partial t}+\frac{1}{h}a\left(u_{i+1}-u_i\right)=0,
	\label{eq3.3}
\end{equation}
The approximation $u_i$ to the $u\left(x_i\right)$ is calculated by the schemes introduced in section 2.2.

To simplify the analysis, $u(x)$ is assumed to be periodic over the domain $\Omega$, i.e., $u_1=u_{N+1}$. Additionally, the grids are assumed to be uniform, i.e., $h_i=h$. The exact solution is taken as a single Fourier mode, given by
\begin{equation}
	u\left(x,t\right)=\ e^{\mathrm{i}\left(kx-\omega t\right)},
	\label{eq3.4}
\end{equation}
where $k$ is the real wavenumber and $\omega$ is the complex circular frequency.  The exact mean value of this Fourier mode over element  $\Omega_i$ is
\begin{equation}
	{\bar{u}}_i=\frac{1}{h}\int_{x_i}^{x_i+1}e^{\mathrm{i}\left(kx-\omega t\right)}dx=\frac{1}{\mathrm{i}kh}e^{\mathrm{i}\left(kx_i-\omega t\right)}\left(e^{\mathrm{i}kh}-1\right).
	\label{eq3.5}
\end{equation}
The exact $m$-th order derivative $h^mu^{\left(m\right)}\left(x_i\right)$ of this Fourier mode at point $x_i$ is given by
\begin{equation}
	h^mu^{\left(m\right)}\left(x_i\right)=\left(\mathrm{i}kh\right)^me^{\mathrm{i}\left(kx_i-\omega t\right)}.
	\label{eq3.6}
\end{equation}
In Eq. (\ref{eq2.3}), mean values are used to approximate the $m$-th order derivative $h^mu^{\left(m\right)}\left(x_i\right)$. It should be noted that this process inevitably introduces errors. Assume the approximate value of $h^mu_i^{\left(m\right)}$ at point $x_i$ takes the form of
\begin{equation}
	h^mu_i^{\left(m\right)}=G\left(\mathrm{i}kh\right)^m e^{\mathrm{i}\left(k x_{i}-\omega t\right)},
	\label{eq3.7}
\end{equation}
where $G$ denotes the amplification factor  when $h^m u^{(m) \left(x_i\right)}$ is approximated using the mean values ${\bar{u}}_i$.

For convenience, a scaled wavenumber $\beta=kh$ and a scaled coordinate ${\widetilde{x}}_j=\frac{x_j-x_i}{h}$ are defined. Eqs. (\ref{eq3.5}) and (\ref{eq3.7}) can thus be rewritten as
\begin{equation}
	{\bar{u}}_j=\frac{1}{\mathrm{i}\beta}e^{\mathrm{i}\left(kx_i-\omega t\right)}\left(e^{\mathrm{i}\beta}-1\right)e^{\mathrm{i}\beta{\widetilde{x}}_j},\ \  
	h^mu_j^{\left(m\right)}=G\left(\mathrm{i}\beta\right)^m e^{\mathrm{i}\left(k x_{i}-\omega t\right)
	}e^{\mathrm{i}\beta{\widetilde{x}}_{q_j}}.
	\label{eq3.7.1}
\end{equation}
For uniform grids, ${\widetilde{x}}_j=j-i$. The range of the scaled wavenumber $\beta$ is $\left[0,\pi\right]$. The variable vector $\bm{f}$ is divided into two components: $\bm{f}_1$ and $\bm{f}_2$, i.e.,
\begin{equation}
	\begin{aligned}
		& \bm{f}_1=\left[{\bar{u}}_1,{\bar{u}}_2,\ \ldots,{\bar{u}}_s\right]_{1\times s},\\
		& \bm{f}_2=\left[h^mu_{q_1}^{\left(m\right)},h^mu_{q_2}^{\left(m\right)},\ldots,h^mu_{q_{n_m}}^{\left(m\right)}\right]_{1\times\left(n-s\right)}.
		\label{eq3.8}
	\end{aligned}
\end{equation}
Here, $\bm{f}_1$ denotes the mean value vector, and $\bm{f}_2$ denotes the $m$-th order derivative vector. 

Substituting  Eq. (\ref{eq3.7.1}) into Eq. (\ref{eq3.8}), we can obtain the expressions for $\bm{f}_1$ and  $\bm{f}_2$ in a single Fourier mode. Define
\begin{equation}
	\bm{g}_1=\left[e^{\mathrm{i}\beta{\widetilde{x}}_1},e^{\mathrm{i}\beta{\widetilde{x}}_2},\ \ldots,e^{\mathrm{i}\beta{\widetilde{x}}_s}\right]_{1\times s},\ \ \bm{g}_2=\left[e^{\mathrm{i}\beta{\widetilde{x}}_{q_1}},e^{\mathrm{i}\beta{\widetilde{x}}_{q_2}},\ldots,e^{\mathrm{i}\beta{\widetilde{x}}_{q_{n_m}}}\right]_{1\times\left(n-s\right)}.
	\label{eq3.9}
\end{equation}
The expressions of $\bm{f}_1$ and $\bm{f}_2$ can be written as follows,
\begin{equation}
	\bm{f}_1=\frac{1}{\mathrm{i}\beta}e^{\mathrm{i}\left(kx_i-\omega t\right)}\left(e^{\mathrm{i}\beta}-1\right)\bm{g}_1,\ \ \bm{f}_2=G\left(\mathrm{i}\beta\right)^me^{\mathrm{i}\left(kx_i-\omega t\right)}\bm{g}_2.
	\label{eq3.10}
\end{equation}
Similarly, the coefficient vector $\bm{c}$ is divided into two components: $\bm{c}_1$ and $\bm{c}_2$. From Eq. (\ref{eq2.11}), we can get
\begin{equation} 
	\bm{c}_1=\mathbb{C}_1\bm{\eta},\ \ \bm{c}_2=\mathbb{C}_2\bm{\eta},
	\label{eq3.11}
\end{equation}
where $\bm{c}_1$ is the coefficient vector corresponding to $\bm{f}_1$ and $\bm{c}_2$ is the coefficient  vector corresponding to $\bm{f}_2$.
Matrices $\mathbb{C}_1$ and $\mathbb{C}_2$ consist of the first $s$ rows and the last $n-s$ rows of $\mathbb{C}$, respectively.

Substituting Eqs. (\ref{eq3.9})$\sim$(\ref{eq3.11}) into Eq. (\ref{eq2.3}), we can get the expression of the amplification factor $G$, i.e.,
\begin{equation}
	G=-\frac{1}{\left(\mathrm{i}\beta\right)^{m+1}}\left(e^{\mathrm{i}\beta}-1\right)\frac{\bm{g}_1\mathbb{C}_1\bm{\eta}}{\bm{g}_2\mathbb{C}_2\bm{\eta}}.
	\label{eq3.12}
\end{equation}
The amplification factor $G$ quantifies the Fourier mode errors of the compact finite volume scheme when approximating $h^mu_{q_i}^{\left(m\right)}$.  $G$ is a complex number and it exhibits the following property: the closer its magnitude is to 1, the more accurate the approximation of the derivative by the compact finite volume method.

When solving the linear convection equation via the compact finite volume method, only mean values and function values need to be considered. Consequently, only the case $m=0$ is analyzed here. Substituting Eqs.(\ref{eq3.5}), (\ref{eq3.7})  and (\ref{eq3.12}), into Eq. (\ref{eq3.3})  yields  $\frac{\omega}{ak}=G$, where $G$ is the amplification factor (as defined earlier) and  $m=0$, i.e.,
\begin{equation}
	\frac{\omega}{ak}=G=-\frac{1}{\mathrm{i}\beta}\left(e^{\mathrm{i}\beta}-1\right)\frac{\bm{g}_1\mathbb{C}_1\bm{\eta}}{\bm{g}_2\mathbb{C}_2\bm{\eta}}.
	\label{eq3.13}
\end{equation}
From Eq. (\ref{eq3.13}), the circular frequency $\omega$ of the semi-discrete scheme can be expressed as a function of the scaled wavenumber $\beta$ and the parameter vector $\bm{\eta}$, i.e.,
\begin{equation}
	\omega=ak\cdot G\left(\beta,\bm{\eta}\right).
	\label{eq3.14}
\end{equation}
The Fourier mode $u\left(x,t\right)=\ e^{\mathrm{i}\left(kx-\omega t\right)}$ can be rewritten as 
\begin{equation}
	u\left(x,t\right)=\ e^{\omega_it}e^{\mathrm{i}\left(kx-\omega_rt\right)}.
	\label{eq3.15}
\end{equation}
Here, $\omega_i$ and $\omega_r$  denote the imaginary and real parts of the circular frequency $\omega$, respectively.

The value of $\omega_i$ characterizes the dissipation and stability of the semi-discrete scheme. For a specific scaled wavenumber $\beta$, if $\omega_i>0$, the amplitude of the Fourier mode increases over time. In this case, the scheme is unstable. Conversely, if $\omega_i<0$ for all $\beta \in (0,\pi)$, the amplitude of the Fourier mode decreases. This prevents error growth over time, ensuring the scheme is stable. For $\omega_i<0$, the larger the magnitude $\|\omega_i\|$, the stronger the scheme’s dissipation. However, excessive numerical dissipation smooths out the details and features of the physical phenomenon.  Thus, $\omega_i$ is expected to be negative but close to zero, balancing stability and low numerical dissipation.

The value of $\omega_r$ characterizes the dispersion of the semi-discrete scheme. The exact value of $\omega_r$ is  $\omega_r=ak$. If $\frac{\omega_r}{ak}>1$, the numerical wave propagates faster than the exact wave, and the scheme is termed a ‘fast scheme’. Conversely, if $\frac{\omega_r}{ak}<1$, the numerical wave propagates slower than the exact wave, and the scheme is termed a ‘slow scheme’. In some numerical simulations, it is necessary to distinguish between fast and slow schemes. For example, NND methods \cite{ZHANG1991193} use a fast scheme upstream of a discontinuity and a slow scheme downstream to capture discontinuities and suppress oscillations. Similarly, $\omega_r$ is expected to be close to its exact value to minimize dispersion.

Different values of $\bm{\eta}$ yield different compact schemes, i.e., different coefficient vectors $\bm{c}$ from Eq. (\ref{eq2.11}). These schemes possess the same accuracy but exhibit distinct dispersion and dissipation characteristics.
This enables control over scheme dispersion and dissipation by adjusting $\bm{\eta}$, facilitating the development of more accurate and robust schemes.

From Eqs. (\ref{eq2.11}) and (\ref{eq2.12}), the coefficient vector $\bm{c}$ has  $n-r-1$  independent free parameters. To control dispersion and dissipation independently, at least two independent free parameters are required for $\bm{c}$. This means that Eq. (\ref{eq2.9}) needs to have at least three linearly independent solutions, that is, $n - r \geq 3$.  From the preceding discussion, the rank $r$ of matrix $\mathbb{A}$ in Eq. (\ref{eq2.9}) satisfies $r \leq k+1$, where $k+1$ is the accuracy order of Eq. (\ref{eq2.3}). Thus, for independent control of dispersion and dissipation via $\bm{\eta}$, the basic condition  is $n-\left(k+1\right)\geq 3$. This will be discussed in detail later in the context of Euler equations.

\begin{figure}[!t]
	\centering
	\includegraphics[height=3.2cm]{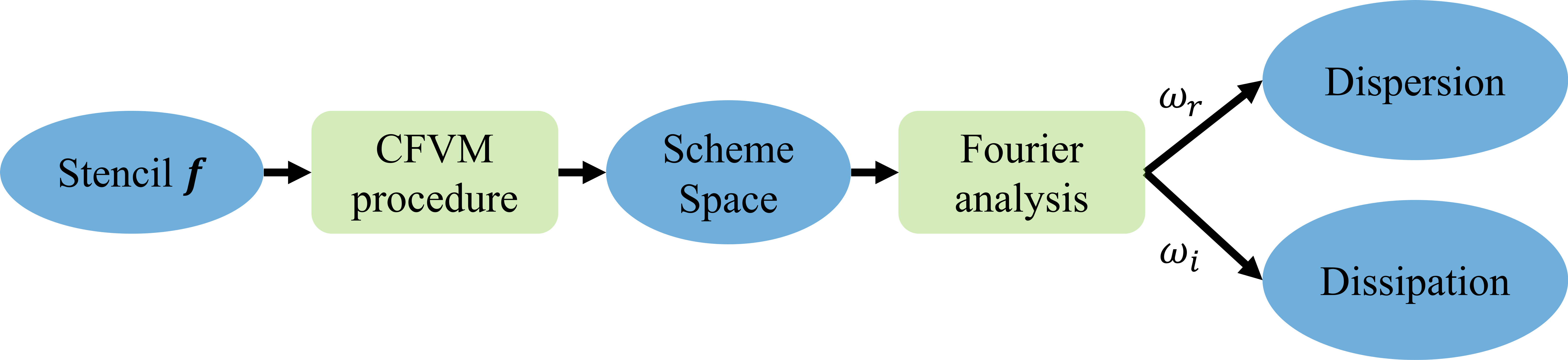}
	\caption{General flow chart for Fourier analysis of compact finite volume schemes.}
	\label{fig_fourier_flowchart}
\end{figure}

Fig. \ref{fig_fourier_flowchart} presents a general flowchart for the Fourier analysis of compact finite volume schemes. For a given stencil variable vector $\bm{f}$, the $(k+1)$-th order scheme space can be obtained via the CFVM procedure described earlier. Fourier analysis is then used to derive the dispersion and dissipation characteristics of the schemes. Eq. (\ref{eq3.14}) provides the theoretical basis for controlling scheme dispersion and dissipation via the parameter vector $\bm{\eta}$. In conjunction with the specific schemes outlined in Section 3, the relationship between scheme dispersion, dissipation, and the control parameter $\eta$ will be explored in detail.

\subsection{Weighted CFVM for Euler equations}
In this section, the concept of nonlinear weighting is incorporated into the proposed compact finite volume method, which is then applied to the Euler equations. The goal of this approach is to suppress numerical oscillations at discontinuities.

The Euler equations in conservation form are given by 
\begin{equation}
	\frac{\partial\bm{U}}{\partial t}+\frac{\partial\bm{F}}{\partial x}=0,
	\label{Eq4.1}
\end{equation}
where $\bm{U}$ is the vector of conservative variables and $\bm{F}$ is the flux vector. For flow problems, $\bm{U}$ and $\bm{F}$ are defined as follows
\begin{equation}
	\bm{U}=\left[\rho,\rho u,\rho E\right]^T,\ \ \bm{F}=\left[\rho u,\rho u^2+p,u\left(\rho E+p\right)\right]^T.
	\label{Eq4.2}
\end{equation}
Here, $\rho$ denotes the fluid density, $u$ is the velocity, and $p$ is the pressure. $E$ is the total energy per unit mass, expressed as 
\begin{equation}
	E=\frac{1}{\gamma-1}\frac{p}{\rho}+\frac{u^2}{2},
	\label{Eq4.3}
\end{equation}
where $\gamma$ denotes the gas adiabatic index. Integrating Eq. (\ref{Eq4.1}) over element $\Omega_i$ and replacing the exact fluxes $\bm{F}$ with numerical fluxes $\hat{\bm{F}}$ yields a semi-discrete finite volume scheme, i.e.,
\begin{equation}
	\frac{\partial{\bar{\bm{U}}}_i}{\partial t}=-\frac{1}{h_i}\left({\hat{\bm{F}}}_{i+1}-{\hat{\bm{F}}}_i\right),
	\label{Eq4.4}
\end{equation}
where ${\hat{\bm{F}}}_i=\hat{\bm{F}}\left(\bm{U}_i\right)$. 
Using the element mean values $\bar{\bm{U}}_i$ , the conserved quantities $\bm{U}_i$ at each point $x_i$ can be reconstructed via Eq. (\ref{eq2.13}). Then, the numerical fluxes $\hat{F}_i$ can be calculated.

For compressible flows, a flux-splitting method is required to compute the fluxes, i.e.,
\begin{equation}
	{\hat{\bm{F}}}_i=\hat{\bm{F}}\left(\bm{U}_i^L,\bm{U}_i^R\right).
	\label{Eq4.5}
\end{equation}
Here $\bm{U}_i^L$ and $\bm{U}_i^R$ are the conserved quantities to the left and right of  point $x_i$, obtained via left-upwind and right-upwind compact schemes, respectively. In this paper, the Lax-Friedrichs and Steger-Warming flux-splitting methods are employed to calculate the numerical fluxes.

When discontinuities exist in the solution, additional measures are required to suppress oscillations at the discontinuities. In this paper, we adopt a strategy analogous to WENO to obtain  oscillation-suppressing  conserved quantities  $\bm{U}_i$. 
The core concept is to use nonlinear weighted compact finite volume schemes, termed the Weighted Compact Finite Volume (WCFV) method. The detailed construction of this method is as follows.

The variable vectors used to solve for the conserved quantities $\bm{U}_i$ are denoted by $\bm{f}_1,\bm{f}_2,\ldots,\bm{f}_n$, and the corresponding coefficient vectors are given by $\bm{c}_1, \bm{c}_2, \ldots, \bm{c}_n$. These coefficient vectors are derived using the method outlined in Subsection 2.1. The conserved quantity derived from variable vector $\bm{f}_r$ and coefficient vector $\bm{c}_r$  is denoted as $\bm{U}_i^r$, with corresponding weights denoted as $\alpha_r,\ r=1,2,\ldots,n$. The weighted conserved quantity is expressed as
\begin{equation}
	\bm{U}_i=\sum_{r=1}^{n}{\alpha_r\bm{U}_i^r}.
	\label{Eq4.6}
\end{equation}
The weights $\alpha_r$ satisfy the condition 
\begin{equation}
	\sum_{r=1}^{n}\alpha_r=1.
	\label{Eq4.7}
\end{equation}
The calculation of weights $\alpha_r$ is analogous to that in the WENO method.

Let the polynomial formed using the element mean values in the variable vector $\bm{f}_r$ be $p_r\left(x\right)$, whose order is assumed as $k-1$. Then, the smooth indicator $\beta_r$ corresponding to $\bm{f}_r$ can be calculated as follows,
\begin{equation}
	\beta_r=\sum_{\ell=1}^{k-1}{\int_{x_{i-1/2}}^{x_{i+1/2}}h^{2\ell-1}\left(\frac{\partial^\ell p_r\left(x\right)}{\left(\partial x\right)^\ell}\right)^2dx}.
	\label{Eq4.8}
\end{equation}
At discontinuities, $\beta_r \sim O\left(1\right)$. In smooth regions, $\beta_r \sim O\left(h^k\right)$. The weight $\alpha_r$ is given by 
\begin{equation}
	 \alpha_r=\frac{\xi_r}{\sum_i^n \xi_i}, \ \ \ 
	 \xi_r=\frac{d_r}{\left(\varepsilon_r+\beta_r\right)^2},
	\label{Eq4.9}
\end{equation}
where $\varepsilon_r$ is a small positive constant (typically $\varepsilon_r={10}^{-6}$) to prevent the denominator from being zero. The constant $d_r$ is a positive value, representing the reference weight of stencil $\bm{f}_r$. The value of $d_r$ indicates the contribution of stencil $\bm{f}_r$ to the conserved quantity $\bm{U}_i$.

From Eq. (\ref{eq2.13}), $\bm{U}_i^r$ is solved via the relaxation iteration method, i.e., 
\begin{equation}
	\left(\bm{U}_i^r\right)^{j+1}=\omega\left(\bm{f}\cdot\bm{c}\right)^j+\left(\bm{U}_i^r\right)^j.
	\label{Eq4.10}
\end{equation}
where $\omega$ is the relaxation coefficient with $0<\omega<2$.
The superscript $j$ denotes the value at the $j$-th iteration step.

Substituting Eqs. (\ref{Eq4.7}) and (\ref{Eq4.10}) into Eq. (\ref{Eq4.6}), we can obtain the iterative solution of the conserved quantity $\bm{U}_i$, i.e.,
\begin{equation}
	\left(\bm{U}_i\right)^{j+1}=\omega\sum_{r}\left(\alpha_r\bm{f}_r\cdot\bm{c}_r\right)^j+\left(\bm{U}_i\right)^j.
	\label{Eq4.11}
\end{equation}
By selecting the variable vectors $\bm{f}_r^L$ and $\bm{f}_r^R$  (to the left and right of point $x_i$), $\bm{U}_i^L$ and $\bm{U}_i^R$ (the conserved quantities on either side of $x_i$) can be solved via Eq. (\ref{Eq4.5}), respectively. The numerical flux $\hat{\bm{ F}}_i$ is then computed using Eq. (\ref{Eq4.5}).

\begin{figure}[!t]
	\centering
	\includegraphics[height=6.5cm]{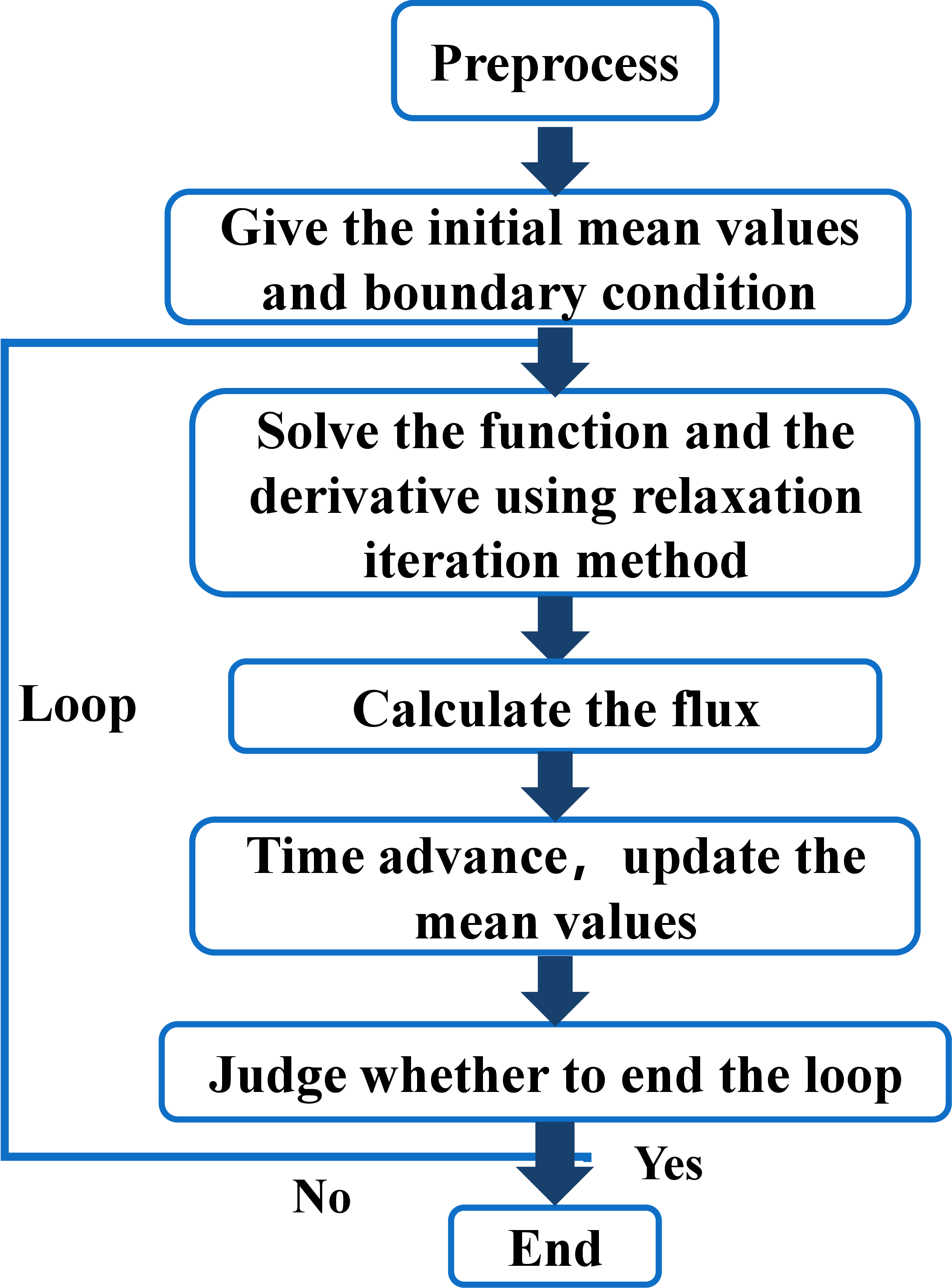}
	\caption{General procedure for solving Euler equations.}
	\label{fig_flowchart}
\end{figure}

The time discretization schemes employed in this paper are the third-order TVD (Total Variation Diminishing) Runge-Kutta method and the fourth-order Runge-Kutta method. Consider a general differential equation of the form, i.e.,
\begin{equation}
	\frac{\partial\bm{U}}{\partial t}=Q\left(\bm{U}\right).
	\label{Eq4.12}
\end{equation}
The third-order TVD Runge-Kutta method \cite{SHU1988439} employed in this paper is
\begin{equation}
	\begin{aligned}
		&\bm{U}^{\left(1\right)}=\bm{U}^n+\Delta t\cdot Q\left(\bm{U}^n\right),\\
		&\bm{U}^{\left(2\right)}=\frac{3}{4}\bm{U}^n+\frac{1}{4}\bm{U}^{\left(1\right)}+\frac{\Delta t}{4}\cdot Q\left(\bm{U}^{\left(1\right)}\right),\\
		&\bm{U}^{n+1}=\frac{1}{3}\bm{U}^n+\frac{2}{3}\bm{U}^{\left(2\right)}+\frac{2\Delta t}{3}\cdot Q\left(\bm{U}^{\left(2\right)}\right).
		\label{Eq4.13}
	\end{aligned}
\end{equation}
The fourth-order  Runge-Kutta method employed  in this paper is
\begin{equation}
	\begin{aligned}
		&\bm{U}^{\left(1\right)}=\bm{U}^n+\frac{\Delta t}{2}\cdot Q\left(\bm{U}^n\right),\\
		&\bm{U}^{\left(2\right)}=\bm{U}^n+\frac{\Delta t}{2}\cdot Q\left(\bm{U}^{\left(1\right)}\right),\\
		&\bm{U}^{\left(3\right)}=\bm{U}^n+\Delta t\cdot Q\left(\bm{U}^{\left(2\right)}\right),\\
		&\bm{U}^{n+1}=\frac{1}{3}\left(-\bm{U}^n+\bm{U}^{\left(1\right)}+2\bm{U}^{\left(2\right)}+\bm{U}^{\left(3\right)}\right)+\frac{\Delta t}{6}\cdot Q\left(\bm{U}^{\left(3\right)}\right).
		\label{Eq4.14}
	\end{aligned}
\end{equation}
Fig. \ref{fig_flowchart} presents a general procedure for solving differential equations. In the initial stage, variable vectors $\bm{f}$ are selected and the corresponding coefficient vectors $\bm{c}$ are computed, as outlined in Subsection 2.1. Subsequently, the conserved quantities $\bm{U}_i$ are computed using Eq. (\ref{Eq4.11}) or Eq. (\ref{eq2.13}). The numerical fluxes $\hat{\bm{ F}}_i$ are then computed using Eq. (\ref{Eq4.5}). Thereafter, the element mean values $\bar{\bm{U}}_i$ are updated via the time integration schemes presented in Eqs. (\ref{Eq4.13}) and (\ref{Eq4.14}). The iteration loop is terminated when the difference in conserved quantity mean values between consecutive time steps is less than the specified tolerance, or when the maximum number of iterations is reached. Otherwise, the aforementioned iteration loop is repeated.

\subsection{Remarks}
In this paper, we introduce a novel method for constructing high-accuracy compact finite volume schemes from the null space perspective. The key to constructing compact finite volume schemes lies in expressing the approximations of function values and derivatives as a linear combination of the given mean values of a set of elements. Specifically, mean values of stencil elements, along with function values and derivatives of various orders at stencil points, are used to construct compact schemes with $(k+1)$-th order accuracy, as shown in Eq. (\ref{eq2.3}). To ensure the existence of solutions for the scheme coefficients, the number of selected variables must be at least the minimum required to achieve the desired accuracy.

Unlike classical compact schemes, where the coefficient vector $\bm{c}$ is determined by matching Taylor series coefficients of different orders, an equivalent construction method is proposed herein. Specifically, a compact scheme achieves $k+1$-th order accuracy means that Eq. (\ref{eq2.3}) holds exactly when the spatial distribution is  any polynomial of degree $k$.  At this point, an underdetermined homogeneous linear system for $\bm{c}$ can be derived, as shown in Eq. (\ref{eq2.9}). Through the above procedure, the problem of constructing compact schemes is effectively transformed into solving the null space of the underdetermined homogeneous linear system. The latter problem has well-established and straightforward solution methods, enabling the convenient construction of a family of compact schemes. The proofs for Eqs. (\ref{eq2.14})$\sim$ (\ref{eq2.16}) demonstrate that following the construction procedure outlined in Eqs. (\ref{eq2.3})$\sim$(\ref{eq2.9}) yields a compact scheme with a truncation error of $O(h^{k+1})$. Numerical validation of this method will be provided in Section 4.

Our method offers  a new perspective on the construction of compact schemes.  For a given stencil, the null space consists of all schemes with accuracy of $(k+1)$-th order or higher, and is termed the ‘scheme space’ for that stencil. The scheme space is a linear vector space. Any linear combination of vectors within the scheme space remains in this space.

These compact schemes can be expressed as a linear combination of the basis vectors $\bm{c}_1,\ldots,\bm{c}_{n-r}$, as shown in Eq. (\ref{eq2.11}). Each value of the parameter vector $\bm{\eta}$ corresponds to a compact scheme with the same or higher accuracy but distinct dispersion and dissipation characteristics. Furthermore, Fourier analysis enables the derivation of the dispersion and dissipation characteristics of these schemes, as presented in Eq. (\ref{eq3.13}).The sets of all dispersion and dissipation characteristics of these schemes are termed the ‘dispersion space’ and ‘dissipation space’, respectively. By selecting different values of $\bm{\eta}$ in the scheme space, we can effectively adjust the dispersion and dissipation characteristics of the schemes. Compared with previous studies, our method enables flexible control of dispersion and dissipation magnitudes while preserving the accuracy and compactness of the schemes. In the numerical tests in Subsection 4,  specific schemes for controlling dispersion and dissipation are examined.

\begin{figure}[!t]
	\centering
	\begin{subfigure}[b]{0.45\textwidth} 
		\centering
		\includegraphics[height=3cm]{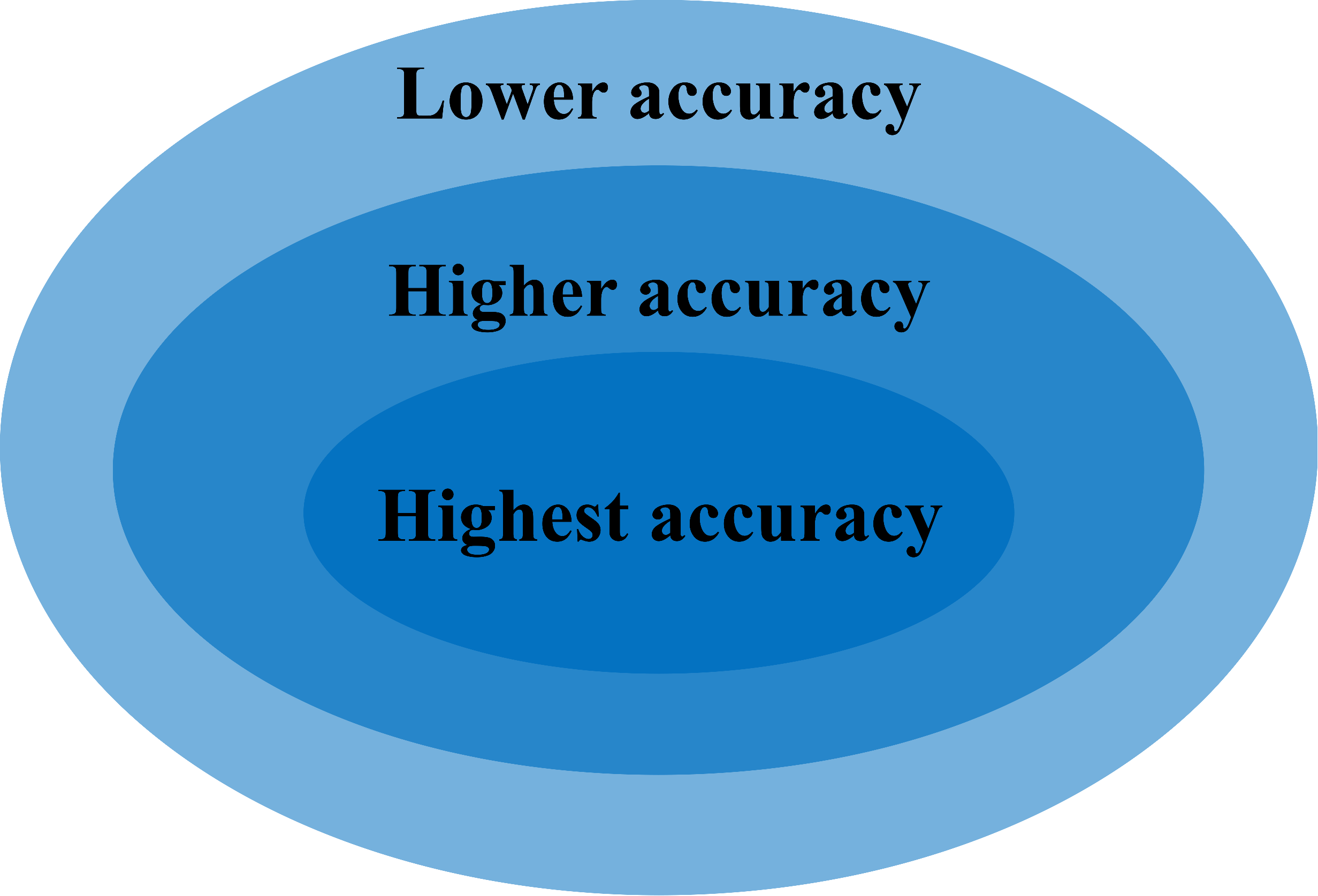}
		\caption{The same stencil} 
		\label{fig2.a} 
	\end{subfigure}
	\hspace{0.2in} 
	\begin{subfigure}[b]{0.45\textwidth}
		\centering
		\includegraphics[height=3cm]{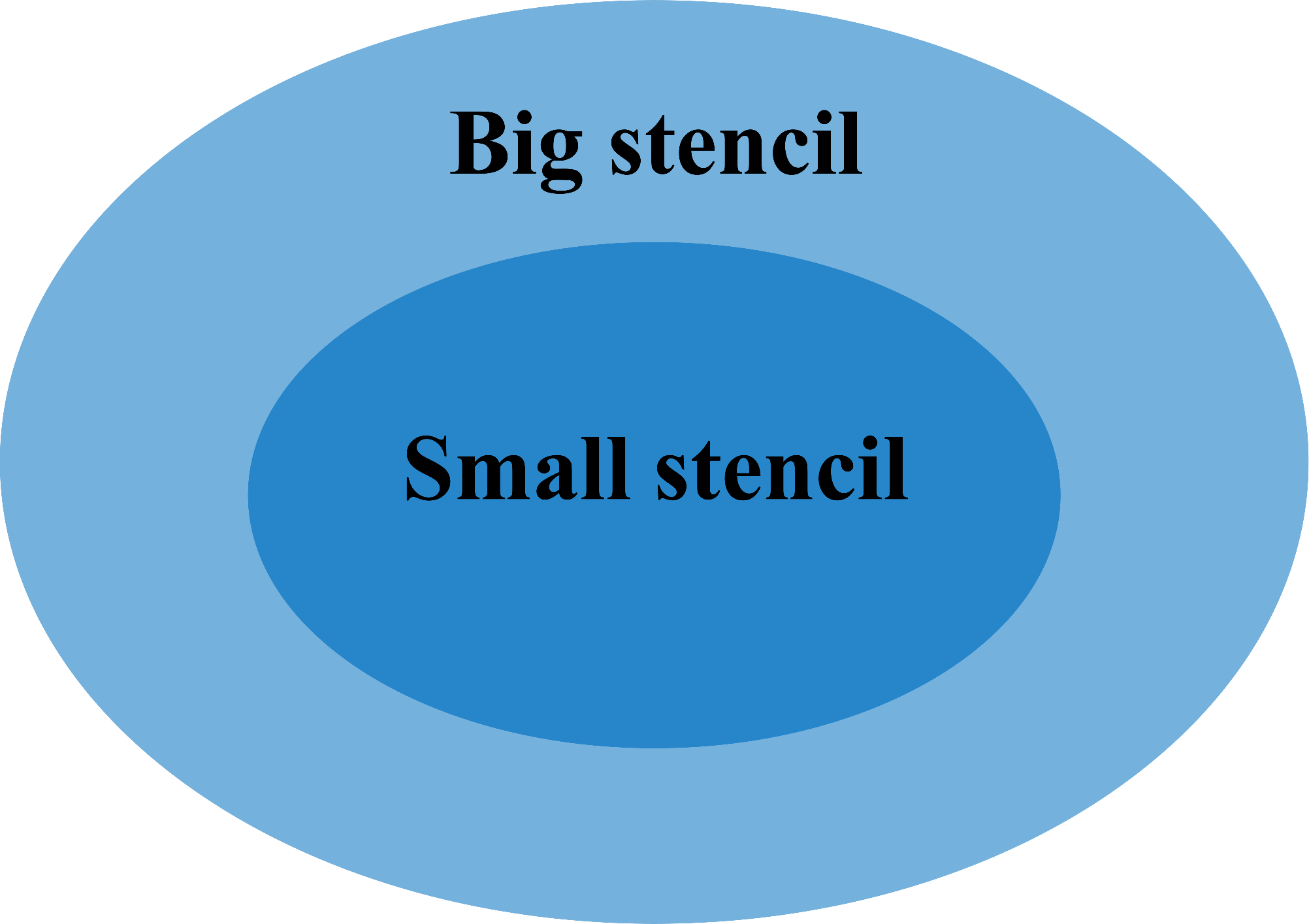}
		\caption{The same accuracy} 
		\label{fig2.b}
	\end{subfigure}
	\caption{Venn diagrams of  the   scheme space, dispersion space and dissipation space for different cases.}
	\label{fig2} 
\end{figure}

The scheme space, dispersion space, and dissipation space exhibit the following  characteristics. First, for uniform grids, these spaces are position-invariant; for non-uniform grids, however, they vary with position. Moreover, as shown in Fig. \ref{fig2.a}, for a given stencil, the spaces (including the scheme space, dispersion space, and dissipation space) of lower accuracy are supersets of those of higher accuracy. Notably, there exists an upper limit on the achievable accuracy of schemes for this stencil. When the desired accuracy exceeds this upper limit, the dimensionality of the scheme space reduces to zero. At this point, if we want  to further improve the scheme accuracy, it is necessary to expand the stencil, i.e., to increase the number of variables in the vector $\bm{f}$.  Additionally, as shown in Fig. \ref{fig2.b}, for the same accuracy, if the variable set of a larger stencil includes that of a smaller stencil, the spaces of the larger stencil will contain those of the smaller stencil.

This paper presents a detailed introduction of the null space-based  compact finite volume method, with particular emphasis on its fundamental properties. While the current discussion focuses on the comprehensive derivation and validation of one-dimensional formulations, it is emphasized that the proposed method is inherently extensible to multidimensional unstructured grids. Compared with classical compact methods, the proposed method addresses the challenges of unpredictable topology in unstructured grids and the complexity of Taylor expansion across various geometries. This enhances the method’s generalization capability while preserving its simplicity. Similar to the construction of one-dimensional schemes, the problem of constructing compact schemes on multidimensional unstructured grids is ultimately transformed into solving the null space of an underdetermined homogeneous linear system.

Here we briefly describe the construction process of the compact finite volume scheme for two-dimensional unstructured grids.

The constructed schemes still follow the form of Eq. (\ref{eq2.3}), where the variable vector $\bm{f}$ consists of mean values of stencil elements and $m$-th order partial derivatives or normal derivatives at stencil points. The total number of variables is still denoted by $n$. The basis functions of 2D $k$-th order polynomials are given by $\bm{\varphi}\left(x\right)=\left[1,x,y,x^2,xy,y^2,\ldots,x^{k-j}y^j\right]^T$.  The number of these basis functions is $n_b=\frac{(k+1)(k+2)}{2}$.   Substituting the basis function vectors $\bm{\varphi}$ into Eq. (\ref{eq2.3}) yields an underdetermined homogeneous linear system, i.e.,
\begin{equation}
	\mathbb{A}\bm{c}=0. 
	\label{Eq4.17}
\end{equation}
The $(i,j)$-th element of matrix $\mathbb{A}$ corresponds to the value of the $i$-th variable in vector $\bm{f}$  when the function is the $j$-th basis function in $\bm{\varphi}$. To ensure the existence of non-trivial solutions to Eq. (\ref{Eq4.17}), the number of variables $n$ must be greater than the number of basis functions $n_b$ (i.e., $n> n_b$). Subsequently, the general solution to this underdetermined system can be derived, with an expression analogous to that of Eq. (\ref{eq2.12}). At this point, the  $k+1$-th order scheme space for a given stencil $\bm{f}$ on unstructured grids is determined. All schemes within this scheme space satisfy the designed accuracy.

In this paper, the variables in Eq. (\ref{eq2.5}) for the compact schemes are the element mean values and the derivative values at specific points.  In fact, the selection of variables can be more flexible. The variables in compact schemes can be any operators applied to $u$. The construction process for such schemes is similar, ultimately transforming into the problem of solving the null space.  

This method can be applied to constructing compact schemes for both finite volume and finite difference methods. To achieve this, it is only necessary to exclude the mean values from the variable vector $\bm{f}$ in Eq. (\ref{eq2.5}). Additionally, function values and derivative values at control points can be used to construct compact schemes, enabling the development of high-accuracy compact schemes with smaller stencils.


\section{Specific schemes}
In this section, specific schemes are discussed, categorized into three types based on the selection of unknown variables. These types include schemes for approximating function values, schemes for approximating derivatives, and hybrid schemes. Additionally, a fourth-order compact scheme for 2D unstructured grids is introduced.

For one-dimensional schemes, the following notation is used.  The stencil comprises mean values of elements $S=\left\{\Omega_1,\Omega_2,\ldots,\Omega_s\right\}$ and $m$-th order derivatives of the function at  $Q^m=\left\{q_1^m,q_2^m,\ldots,q_{n_m}^m\right\}$.  A $k+1$-th order compact finite volume scheme constructed using stencil S and points $S$ and $Q^m$  (for $m=0,1,2,\ldots,M$) is denoted as $\text{CFVM-}P(k+1)\text{-}S(s)\text{-}Q^1(n_1)\text{-}Q^2(n_2)\text{-}\ldots\text{-}Q^m(n_m)$.   Here, $s$ is the number of elements in stencil $S$, and $n_m$ denotes the number of points in $Q^m$ at which the $m$-th order derivative is used.

\subsection{Approximation to function}
In this subsection, compact finite volume schemes for approximating function values are introduced. 

The construction of these finite volume schemes involves mean values of elements $S={\Omega_1, \Omega_2, \ldots, \Omega_s}$ and function values at points $Q^0=\left\{q_1, q_2, \ldots, q_{n_0} \right\}$.
The finite volume approximation of function values at points $Q^0$ is expressed as a linear combination of the given mean values of stencil elements $S$, i.e.,
\begin{equation}
	\bm{f}\cdot\bm{c}=R\left(O\left(h^{k+1}\right)\right),
	\label{eq_sub2.1}
\end{equation}
where $\bm{f}=\left[{\bar{u}}_1,{\bar{u}}_2,\ \ldots,{\bar{u}}_s,u_{q_1},u_{q_2},\ldots,u_{q_{n_0}}\right]$ and $\bm{c}=\left[c_1,c_2,c_3,\ldots,c_n\right]^T$ are the vectors of the variables and coefficients,  respectively.

As described in Subsection 2.1, for a given stencil $\bm{f}$, the construction of these schemes involves finding the general solution for the coefficient vector $\bm{c}$, as presented in Eq. (\ref{eq2.11}). As discussed earlier, to control the dispersion and dissipation of the schemes independently, we consider schemes that satisfy the condition $s+n_0-\left(k+1\right)\geq3$. Here, a fourth-order accurate compact scheme is presented, namely the $\text{CFVM-}P(4)\text{-}S(4)\text{-}Q^0(3)$ scheme.

\begin{figure}[!t]
	\centering
	\includegraphics[width=7.5cm]{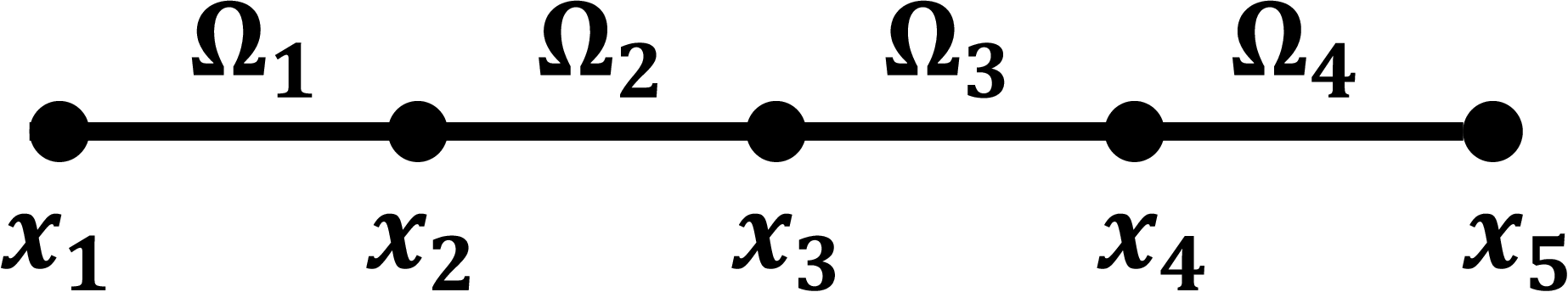}
	\caption{Grid stencil in one-dimension.}
	\label{fig_grid_stencil}
\end{figure}

This scheme uses variables that include mean values of four elements and function values at three points, and is constructed to be fourth-order accurate. Fig. \ref{fig_grid_stencil} presents the grid stencil for one-dimensional compact finite volume schemes.
The elements in the stencil are
\begin{equation}
	S=\left\{\Omega_1,\Omega_2,\Omega_3,\Omega_4\right\}.
	\label{eq_sub2.2}
\end{equation}
Three points are selected to construct the scheme, with three possible choices, i.e.,
\begin{equation}
	Q_1^0=\left\{x_1,x_2,x_3\right\},\ \ Q_2^0=\left\{x_2,x_3,x_4\right\},\ \ Q_3^0=\left\{x_3,x_4,x_5\right\}.
	\label{eq_sub2.3}
\end{equation}
First, the symmetric stencil is considered, consisting of $S$ and $Q_2^0$. The variable and coefficient vectors are given by
\begin{equation}
	\bm{c}=\left[c_1,c_2,\ldots,c_7\right]^T,\ \ \bm{f}=\left[{\bar{u}}_1,\ {\bar{u}}_2,\ {\bar{u}}_3,\ {\bar{u}}_4,u_2,\ u_3,\ u_4\right].
	\label{eq_sub2.4}
\end{equation}
The basis functions of a cubic polynomial are $\bm{\varphi}\left(x\right)=\left[1,x,x^2,x^3\right]^T$. As discussed in Subsection 2.1, when constructing this fourth-order scheme, Eq. (\ref{eq2.3}) holds exactly for all basis functions of a cubic polynomial. From Eqs. (\ref{eq2.10}) and (\ref{eq2.10.new}), the expression for matrix A can be derived, i.e.,
\begin{equation}
	\mathbb{A}=
	\left[
	\begin{matrix}1&1&1&1&1&1&1\\
		\frac{1}{{2h}_1}\left(x_2^2-x_1^2\right)&\frac{1}{2h_2}\left(x_3^2-x_2^2\right)&\frac{1}{{2h}_3}\left(x_4^2-x_3^2\right)&\frac{1}{2h_4}\left(x_5^2-x_4^2\right)&x_2&x_3&x_4\\
		\frac{1}{{3h}_1}\left(x_2^3-x_1^3\right)&\frac{1}{3h_2}\left(x_3^3-x_2^3\right)&\frac{1}{3h_3}\left(x_4^3-x_3^3\right)&\frac{1}{{3h}_4}\left(x_5^3-x_4^3\right)&x_2^2&x_3^2&x_4^2\\
		\frac{1}{{4h}_1}\left(x_2^4-x_1^4\right)&\frac{1}{4h_2}\left(x_3^4-x_2^4\right)&\frac{1}{4h_3}\left(x_4^4-x_3^4\right)&\frac{1}{{4h}_4}\left(x_5^4-x_4^4\right)&x_2^3&x_3^3&x_4^3\\
		\label{eq_sub2.5}
	\end{matrix}
	\right],
\end{equation}
where $h_i=x_{i+1}-x_i$ is the length of element $\Omega_i$.  The general solution for the coefficient vector $\bm{c}$ of this underdetermined homogeneous linear system is then derived as
\begin{equation}
	\bm{c}=\eta_1\bm{c}_1+\eta_2\bm{c}_2+\eta_{3}\bm{c}_{3}=\mathbb{C}\bm{\eta}.
	\label{eq_sub2.6}
\end{equation}
Here, $\mathbb{C}$ is a $7\times3$ matrix and $\bm{\eta}=\left[\eta_1,\eta_2,\eta_3\right]^T$. 

Thus far, the fourth-order accurate scheme space based on stencil $S$ and $Q_2^0$ has been successfully constructed. Within this scheme space, any linear combination of vectors in Eq. (\ref{eq_sub2.6}) is fourth-order accurate. For different forms of cubic polynomial basis function vectors, e.g., $\bm{\varphi}\left(x\right) = \left[1, x-x_0, (x-x_0)^2, (x-x_0)^3 \right]^T$, the scheme spaces described by Eq. (\ref{eq_sub2.6}) remain identical. This indicates that the scheme space depends solely on the selection of the stencil and the relative positions of stencil points and elements. For non-uniform grids, the relative positions between stencil elements and points vary, resulting in position-dependent scheme spaces. For uniform grids, however, the scheme space is position-invariant, meaning that the same matrix $\mathbb{C}$ is obtained at all positions.

Here, the case of uniform grids is discussed, where the matrix $\mathbb{C}$ for the $\text{CFVM-}P(4)\text{-}S(4)\text{-}Q^0_2(3)$ scheme is given by
\begin{equation}
	\mathbb{C}=\frac{1}{12}
	\left(\begin{array}{ccccccc} 
		-3 & -13 & 5 & -1 & 12 & 0 & 0\\ 
		1 & -7 & -7 & 1 & 0 & 12 & 0\\ 
		-1 & 5 & -13 & -3 & 0 & 0 & 12 
	\end{array}\right)^T.
    \label{eq_sub2.6.1}
\end{equation}
Notably, the expression of matrix $\mathbb{C}$  may differ from that in Eq. (\ref{eq_sub2.6.1}), as long as the spaces they span are equivalent. Substituting Eq. (\ref{eq_sub2.6.1})  into Eq. (\ref{eq2.16}) yields the truncation error of this scheme at specific points, i.e., 
\begin{equation}
	R=\sum_{p=4}^{\infty}\bm{f}_p\mathbb{C}\bm{\eta}.
	\label{eq_sub2.6.2}
\end{equation}
Here, $\bm{f}_p$ denotes the variable vector $\bm{f}$ when $u(x)=T_p (x)$. The expressions for $\bm{f}$ and $T_p (x)$  are provided in Eqs. (\ref{eq2.14}) and Eq. (\ref{eq_sub2.4}), respectively.

In this work, only the first three non-zero terms of the truncation error are considered. The $\text{CFVM-}P(4) \text{-} S(4) \text{-} Q^0_2(3)$ scheme can be used to approximate function values at points $x_2$,  $x_3$ and $x_4$.  The truncation error of this scheme at point $x_i$  can be expressed as 
\begin{equation}
	R_{x_i}=\left[\frac{u^{\left(4\right)}\left(x_i\right)}{4!}h^4,\frac{u^{\left(5\right)}\left(x_i\right)}{5!}h^5,\frac{u^{\left(6\right)}\left(x_i\right)}{6!}h^6\right]\mathbb{R}_i\bm{\eta},
	\label{eq_sub2.7}
\end{equation}
where $\mathbb{R}_i$ is a $3\times3$ matrix. For points $x_i$ (where $i=2, 3, 4$), the corresponding truncation error matrices $\mathbb{R}_i$ are
\begin{equation}
	\mathbb{R}_2=\left(\begin{array}{ccc} -\frac{6}{5} & \frac{4}{5} & -\frac{6}{5}\\ -5 & 4 & -7\\ -\frac{120}{7} & \frac{104}{7} & -\frac{204}{7} \end{array}\right),\ \ 
	\mathbb{R}_3=\left(\begin{array}{ccc} -\frac{6}{5} & \frac{4}{5} & -\frac{6}{5}\\ 1 & 0 & -1\\ -\frac{36}{7} & \frac{20}{7} & -\frac{36}{7} \end{array}\right),\ \ 
	\mathbb{R}_4=\left(\begin{array}{ccc} -\frac{6}{5} & \frac{4}{5} & -\frac{6}{5}\\ 7 & -4 & 5\\ -\frac{204}{7} & \frac{104}{7} & -\frac{120}{7} \end{array}\right).
	\label{eq_sub2.7.1}
\end{equation}
From (\ref{eq_sub2.6}) and  Eq. (\ref{eq_sub2.6.1}), a specific scheme can be obtained by selecting appropriate values of the parameter vector $\bm{\eta}$. For instance, when $\bm{\eta}=[0,1,0]$, the central explicit scheme is obtained, i.e.,
\begin{equation}
	u_3=-\frac{1}{12} {\bar{u}}_1 +\frac{7}{12}{\bar{u}}_2  +\frac{7}{12}{\bar{u}}_3  -\frac{1}{12} {\bar{u}}_4.
	\label{eq_schemes_1}
\end{equation}
From Eq. (\ref{eq_sub2.7.1}), when $3\eta_{1}-2\eta_{2}+3\eta_{3}=0$, the fourth-order term in the truncation errr vanishes, allowing the scheme to achieve fifth-order accuracy. In particular, when $\eta_{2} = \frac{1}{3}+\alpha$ and $\eta_{2} = \frac{1}{3}+\alpha$, a fifth-order accurate central dissipative compact scheme is obtained, expressed as
\begin{equation}
	\left(\frac{1}{3}-\alpha\right) u_2 + u_3  + \left(\frac{1}{3}+\alpha\right) u_4 
	=\left(-\frac{\alpha }{6}+\frac{1}{36}\right) {\bar{u}}_1 
	+\left(-\frac{3\,\alpha }{2}+\frac{29}{36}\right) {\bar{u}}_2
	+\left(\frac{3\,\alpha }{2}+\frac{29}{36}\right) {\bar{u}}_3
	+\left(\frac{\alpha }{6}+\frac{1}{36}\right) {\bar{u}}_4.
	\label{eq_schemes_2}
\end{equation}
Furthermore, when $\eta_{1}=\eta_{3}$, the fifth-order term in the truncation error vanishes. Specifically, when $\eta_{1} = \eta_{3} = \frac{1}{3} \eta_{2}$, the scheme achieves sixth-order accuracy. The sixth-order scheme is given by
\begin{equation}
	\frac{1}{3} u_2 + u_3 + \frac{1}{3} u_4
	=	\frac{1}{36} {\bar{u}}_1 + \frac{29}{36} {\bar{u}}_2 +
	\frac{29}{36} {\bar{u}}_3 +	\frac{1}{36} {\bar{u}}_4.
	\label{eq_schemes_3}
\end{equation}
Eqs. (\ref{eq_schemes_1}) $\sim$ (\ref{eq_schemes_3}) represent distinct schemes, which exhibit different dispersion and dissipation characteristics but belong to the same scheme space.

From Eq. (2.32), the expression for the circular frequency $\omega$ can be derived, i.e.,
\begin{equation}
	\frac{\omega}{ak}=G=-\frac{1}{\mathrm{i}\beta}\left(e^{\mathrm{i}\beta}-1\right)\frac{\bm{g}_1\mathbb{C}_1\bm{\eta}}{\bm{g}_2\mathbb{C}_2\bm{\eta}}.
	\label{eq_sub2.8}
\end{equation}
The expressions for $\mathbb{C}_1$ and $\mathbb{C}_2$ are
\begin{equation}
	\mathbb{C}_1=\frac{1}{12}
	\left(\begin{array}{cccc} 
		-3 & -13 & 5   & -1 \\ 
		1  & -7  & -7  & 1 \\ 
		-1 & 5   & -13 & -3 
	\end{array}\right)^T,\ \ \ \ 
	\mathbb{C}_2=
	\left(\begin{array}{ccc} 
		1 & 0 & 0\\ 
		0 & 1 & 0\\ 
		0 & 0 & 1 
	\end{array}\right).
	\label{eq_sub2.9}
\end{equation}
The expressions for $\bm{g}_1$ and $\bm{g}_2$ vary with the target point. For point $x_3$, the expressions for $\bm{g}_1$ and $\bm{g}_2$ are
\begin{equation}
	\bm{g}_{1,x_3}=\left[e^{-2\beta\mathrm{i}},e^{-\beta\mathrm{i}},1,e^{\beta\mathrm{i}}\right],\ \ \ 
	\bm{g}_{2,x_3}=\left[e^{-\beta\mathrm{i}},1,e^{\beta\mathrm{i}}\right].
	\label{eq_sub2.10}
\end{equation}
For point $x_4$, the expressions for $\bm{g}_1$ and $\bm{g}_2$ are
\begin{equation}
	\bm{g}_{1,x_4}=\left[e^{-3\beta\mathrm{i}},e^{-2\beta\mathrm{i}},e^{-\beta\mathrm{i}},1\right],\ \ \ 
	\bm{g}_{2,x_4}=\left[e^{-2\beta\mathrm{i}},e^{-\beta\mathrm{i}},1\right].
	\label{eq_sub2.11}
\end{equation}
Comparing Eqs. (\ref{eq_sub2.10}) and (\ref{eq_sub2.11}) reveals that $\bm{g}_{1,x_3} = e^{\beta\mathrm{i}} \bm{g}_{1,x_4}$ and $\bm{g}_{2,x_3} = e^{\beta\mathrm{i}} \bm{g}_{2,x_4}$. Thus, for the same $\omega$, the expression for $\omega$ in Eq. (\ref{eq_sub2.8}) remains identical, which means that  the dispersion and dissipation characteristics are the same in this case. The specific form of $\omega$ is derived as follows, 
\begin{equation}
	\frac{\omega}{ak}=\frac{1}{6 \beta}\frac{p_1+p_2\cdot i}{q_1-q_2\cdot i},
	\label{eq_sub2.11.1}
\end{equation}
where
\begin{equation}
	\begin{aligned}
	&p_1=2\eta_2\left[-\sin{\left(2\beta\right)}+8\sin{\left(\beta\right)}\right]+4\left(\eta_1+\eta_3\right)\left[\sin{\left(2\beta\right)}+\sin{\left(\beta\right)}\right],\\  
	&p_2=2(\eta_1-\eta_3)\left[-9+8\cos(\beta)+\cos(2\beta) \right], \\   &q_1=\eta_2+\left(\eta_1+\eta_3\right)\cos{\left(\beta\right)}, \ \ 
	q_2=\left(\eta_1-\eta_3\right)\sin{\left(\beta\right)}.
	\label{eq_sub2.11.2}
	\end{aligned}
\end{equation}
From this, the expressions for the imaginary part $\omega_i$ (dissipation) and real part $\omega_r$ (dispersion) of $\omega$ can be derived.  Here, the specific expression for $\omega_i$ is provided, which simplifies to
\begin{equation}
	\frac{\omega_i}{ak}=-\frac{1}{3 \beta}\frac{\left(\cos(\beta)-1\right)^3}{q_1^2+q_2^2}
	\left(\eta_{1}-\eta_{3}\right)
	\left(\eta_{1}-\eta_{2}+\eta_{3}\right).
	\label{eq_sub2.11.3}	
\end{equation}
To ensure the stability of the scheme, $\omega_i<0$ is a necessary condition. For $a>0$, the stability condition of this scheme is derived from Eq. (\ref{eq_sub2.11.3}) as
\begin{equation}
	\left(\eta_{1}-\eta_{3}\right)
	\left(\eta_{1}-\eta_{2}+\eta_{3}\right)\leq 0.
	\label{eq_sub2.11.4}	
\end{equation}

Different values of $\bm{\eta}$ lead to different expressions for $\omega_r$ and $\omega_i$, indicating that the dispersion and dissipation characteristics of the scheme vary.  Taking point $x_3$ as an example, to ensure that the function value at this point can be solved, it is necessary to ensure that the coefficient of $u_3$ in the scheme is non-zero. For simplicity and clarity, we set $\eta_2=1$, so $\bm{\eta} = \left[\eta_1, 1, \eta_3\right]$. At this point, the free parameters are $\eta_1$ and $\eta_3$.

\begin{figure}[!t]
 	\centering
 	\begin{subfigure}[b]{0.45\textwidth} 
 		\centering
 		\includegraphics[height=5cm]{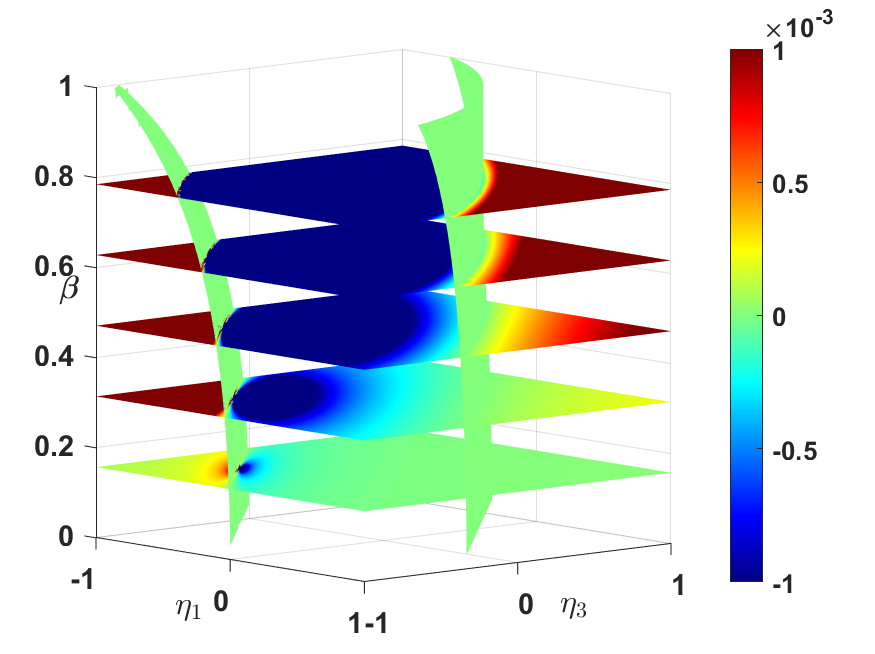}
 		\caption{The dispersion} 
 		\label{fig_dispersion_3D} 
 	\end{subfigure}
 	\hspace{0.2in} 
 	\begin{subfigure}[b]{0.45\textwidth}
 		\centering
 		\includegraphics[height=5cm]{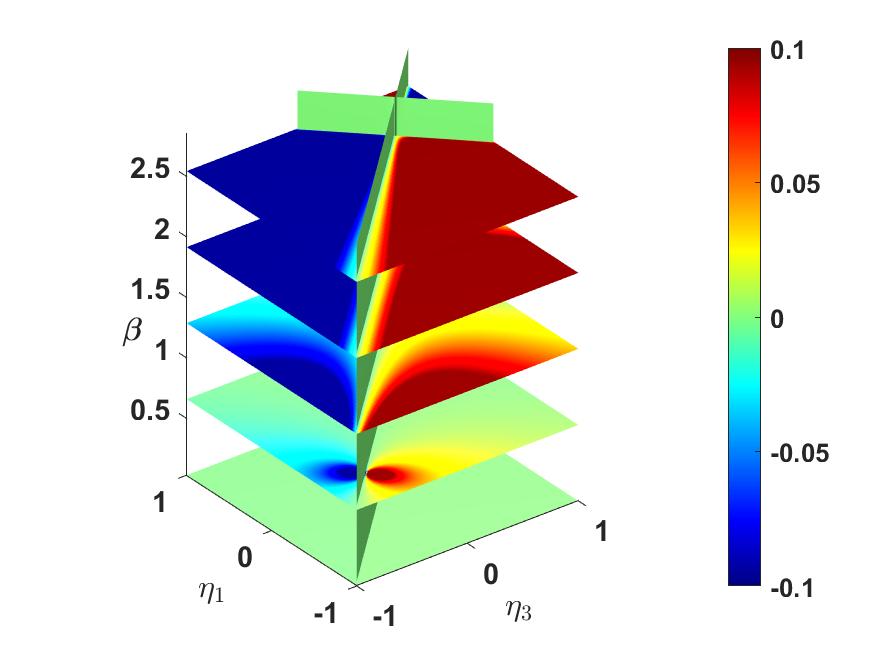}
 		\caption{The dissipation} 
 		\label{fig_dissipation_3D}
 	\end{subfigure}
 	\caption{Dispersion and dissipation characteristics  of the $\text{CFVM-}P(4)\text{-}S(4)\text{-}Q^0(3)$ scheme for approximating function  values at point $x_3$.}
 	\label{fig_fourier_w_3D} 
 \end{figure}

Fig. \ref{fig_fourier_w_3D}  shows how dispersion ($\frac{\omega_r}{ak}\omega_r$) and dissipation ($\omega_i$) vary with parameters $\eta_1$ and $\eta_3$ for different dimensionless wavenumbers $\beta$ when approximating the function value at point $x_3$. Fig. 7 presents dispersion and dissipation curves for specific values of $\bm{\eta}$ (with $\eta_2=1$).

\begin{figure}[!t]
	\centering
	\begin{subfigure}[b]{0.45\textwidth} 
		\centering
		\includegraphics[height=5cm]{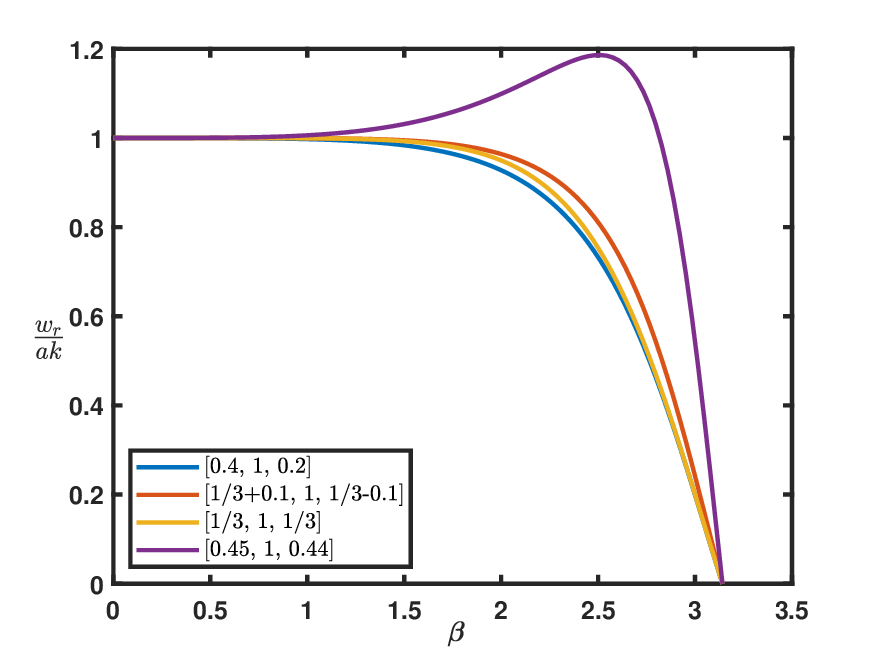}
		\caption{The dispersion} 
		\label{fig_dispersion} 
	\end{subfigure}
	\hspace{0.2in} 
	\begin{subfigure}[b]{0.45\textwidth}
		\centering
		\includegraphics[height=5cm]{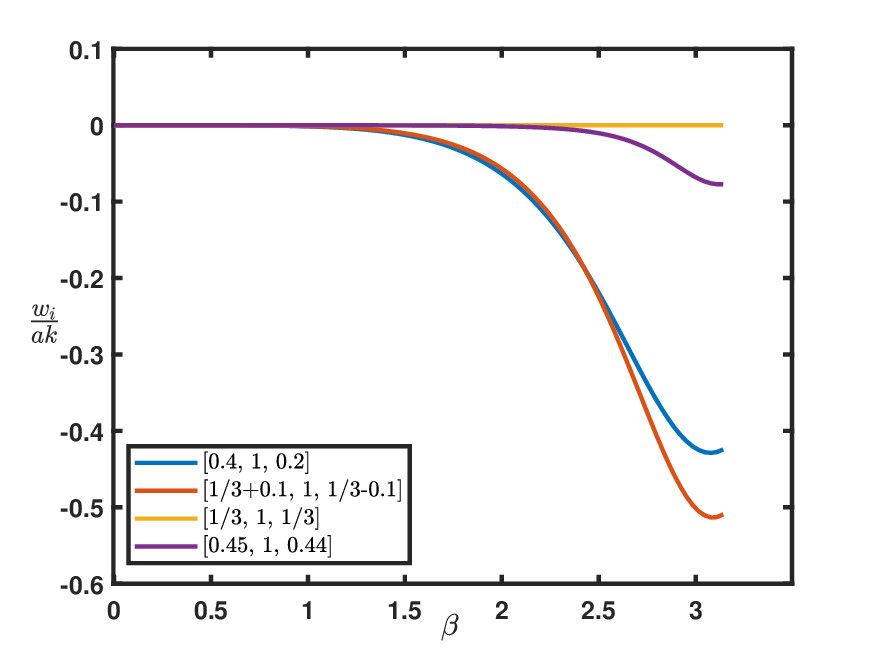}
		\caption{The dissipation} 
		\label{fig_dissipation}
	\end{subfigure}
	\caption{Dispersion and dissipation curves of the $\text{CFVM-}P(4)\text{-}S(4)\text{-}Q^0_2(3)$ schemes for approximating function  values at point $x_3$.}
	\label{fig_fourier_w} 
\end{figure}

Similarly, the dispersion and dissipation characteristics of the $\text{CFVM-}P(4)\text{-}S(4)\text{-}Q^0_2(3)$ scheme can be analyzed for approximating function values at points $x_2$ or $x_4$.
Due to the symmetry of these two points, only the case of point $x_4$ is discussed herein. 

\begin{figure}[!t]
	\centering
	\begin{subfigure}[b]{0.45\textwidth} 
		\centering
		\includegraphics[height=5cm]{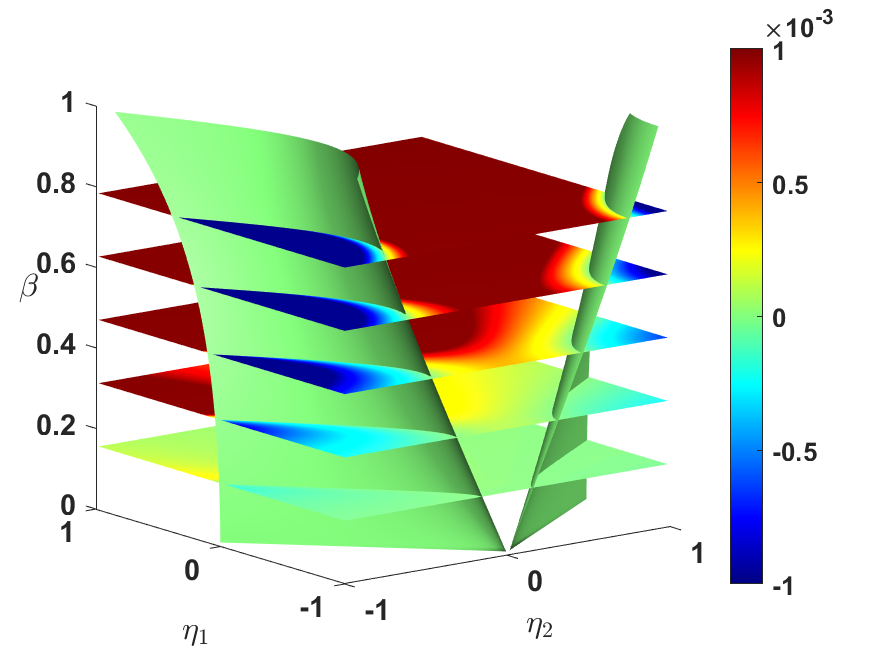}
		\caption{The dispersion} 
		\label{fig_dispersion_3D_x4} 
	\end{subfigure}
	\hspace{0.2in} 
	\begin{subfigure}[b]{0.45\textwidth}
		\centering
		\includegraphics[height=5cm]{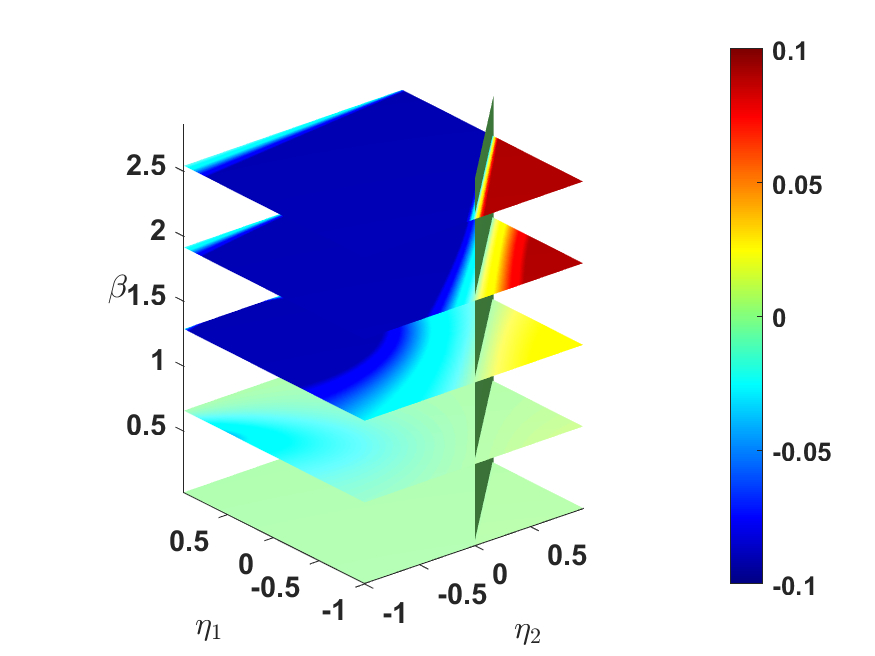}
		\caption{The dissipation} 
		\label{fig_dissipation_3D_x4}
	\end{subfigure}
	\caption{Dispersion and dissipation characteristics of the $\text{CFVM-}P(4)\text{-}S(4)\text{-}Q^0(3)$ scheme for approximating function values at point $x4$.}
	\label{fig_fourier_w_3D_x4} 
\end{figure}

\begin{figure}[!t]
	\centering
	\begin{subfigure}[b]{0.45\textwidth} 
		\centering
		\includegraphics[height=5cm]{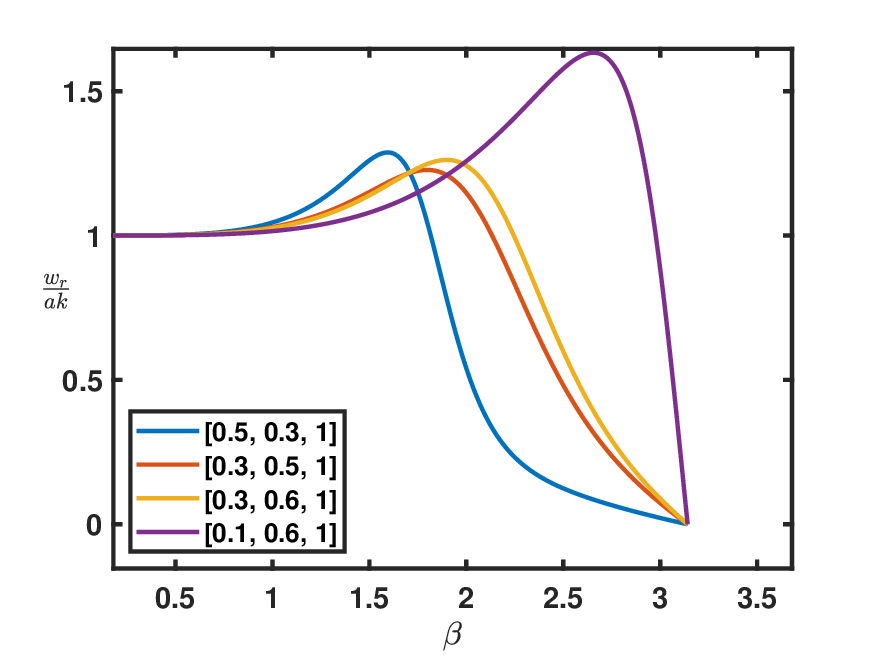}
		\caption{The dispersion} 
		\label{fig_dispersion_x4} 
	\end{subfigure}
	\hspace{0.2in} 
	\begin{subfigure}[b]{0.45\textwidth}
		\centering
		\includegraphics[height=5cm]{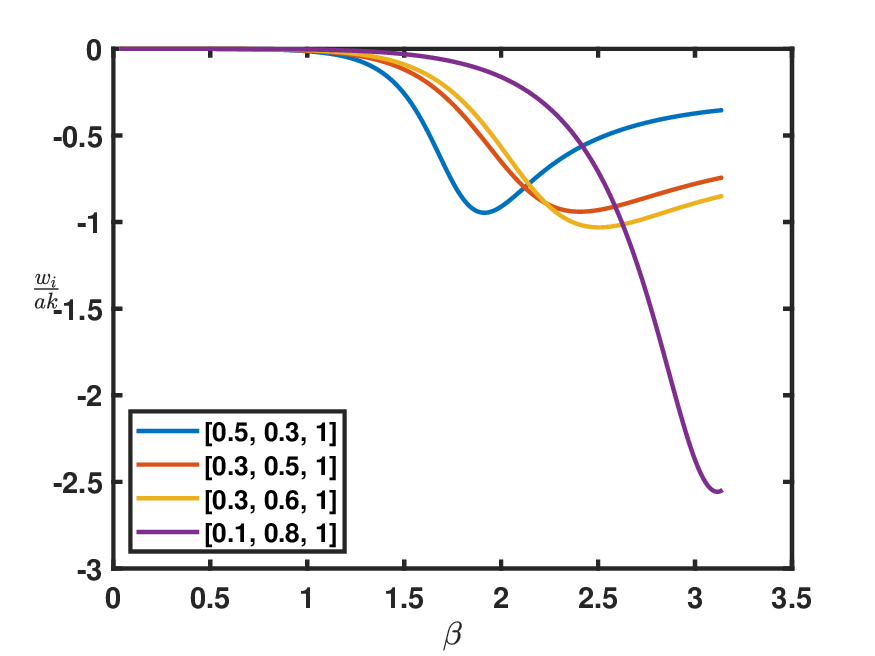}
		\caption{The dissipation}
		\label{fig_dissipation_x4}
	\end{subfigure}
	\caption{Dispersion and dissipation curves of the $\text{CFVM-}P(4)\text{-}S(4)\text{-}Q^0_2(3)$ scheme for approximating function values at point $x4$.}
	\label{fig_fourier_w_x4} 
\end{figure}

To approximate the function value at point $x_4$,, it is necessary to ensure that $\eta_3 \neq 0$. We set $\eta_3=1$, so $\bm{\eta} = \left[\eta_1, \eta_2, 1\right]$. In this case, the free parameters are $\eta_1$ and $\eta_2$. Fig. \ref{fig_fourier_w_3D_x4}  shows the dispersion and dissipation characteristics when approximating the function value at point $x_4$ with different parameters $\eta_1$ and $\eta_2$ for various dimensionless wavenumbers $\beta$. Fig. \ref{fig_fourier_w_x4} presents dispersion and dissipation curves for specific values of $\bm{\eta}$ (with $\eta_3=1$).

A similar analysis can be performed for schemes based on the point stencils $Q_1^0$ or  $Q_3^0$. Since the point stencils $Q_1^0$ or  $Q_3^0$ are symmetric, $Q_1^0$ is used as an example herein. The scheme space follows the form of Eq. (\ref{eq_sub2.6}), where the matrix $\mathbb{C}$ for the $\text{CFVM-}P(4)\text{-}S(4)\text{-}Q^0_1(3)$ scheme is given by
\begin{equation}
	\mathbb{C}=\frac{1}{12}
	\left(\begin{array}{ccccccc} -25 & 23 & -13 & 3 & 12 & 0 & 0\\ -3 & -13 & 5 & -1 & 0 & 12 & 0\\ 1 & -7 & -7 & 1 & 0 & 0 & 12 \end{array}\right)^T.
\end{equation}
The truncation error matrices $Ri$ for points $x_1$, $x_2$ and $x_3$ are
\begin{equation}
	\mathbb{R}_1=\left(\begin{array}{ccc} \frac{29}{5} & \frac{69}{5} & \frac{329}{5}\\ 41 & 20 & 219\\ \frac{1567}{7} & -\frac{15}{7} & \frac{5011}{7} \end{array}\right),\ \ 
	\mathbb{R}_2=\left(\begin{array}{ccc} \frac{19}{5} & -\frac{1}{5} & \frac{79}{5}\\ 17 & -4 & 35\\ \frac{377}{7} & -\frac{113}{7} & \frac{545}{7} \end{array}\right),\ \ 
	\mathbb{R}_3=\left(\begin{array}{ccc} -\frac{51}{5} & -\frac{11}{5} & \frac{9}{5}\\ 23 & 2 & 1\\ -\frac{225}{7} & -\frac{43}{7} & \frac{27}{7} \end{array}\right).
\end{equation}
The expression for circular frequency $\omega$ is given by Eq. (\ref{eq_sub2.8}), i.e.  $\frac{\omega}{ak}=-\frac{1}{\mathrm{i}\beta}\left(e^{\mathrm{i}\beta}-1\right)\frac{\bm{g}_1\mathbb{C}_1\bm{\eta}}{\bm{g}_2\mathbb{C}_2\bm{\eta}}.$
The expressions for $\mathbb{C}_1$ and $\mathbb{C}_2$ in Eq. (\ref{eq_sub2.8}) are
\begin{equation}
	\mathbb{C}_1=\frac{1}{12}
	\left(\begin{array}{cccc} 
		-3 & -13 & 5   & -1 \\ 
		1  & -7  & -7  & 1 \\ 
		-1 & 5   & -13 & -3 
	\end{array}\right)^T,\ \ 
	\mathbb{C}_2=
	\left(\begin{array}{ccc} 
		1 & 0 & 0\\ 
		0 & 1 & 0\\ 
		0 & 0 & 1 
	\end{array}\right).
	\label{eq_sub2.12}
\end{equation}
The expressions for $\bm{g}_1$ and $\bm{g}_2$ are given by
\begin{equation}
	\bm{g}_{1}=\left[1,e^{\beta\mathrm{i}},e^{2\beta\mathrm{i}},e^{3\beta\mathrm{i}}\right],\ \ \ 
	\bm{g}_{2}=\left[1,e^{\beta\mathrm{i}},e^{2\beta\mathrm{i}}\right],.
	\label{eq_sub2.13}
\end{equation}
The dispersion and dissipation characteristics of the scheme can be expressed as a function of $\bm{\eta}$. The analysis of schemes the stencil points $Q^0_1$ is identical to that of schemes using the stencil $Q^0_2$, so it is not repeated herein.

\subsection{Approximation to the first derivative}
In this subsection, compact schemes for approximating first derivatives are introduced.  For these first derivative approximation schemes, the stencil comprises mean values of elements $S={\Omega_1, \Omega_2, \ldots, \Omega_s}$ and first derivatives of the function at points $Q^1 = \left\{q_1^1, q_2^1, \ldots, q_{n_1}^1\right\}$. A $(k+1)$-th order compact finite volume scheme constructed using stencil elements $S$ and stencil points $Q^1$ is denoted as $\text{CFVM-}P(k+1)\text{-}S(s)\text{-}Q^1(n_1)$.

Here, a fourth-order accurate compact scheme is presented, namely the $\text{CFVM-}P(4)\text{-}S(4)\text{-}Q^1(3)$ scheme. The stencil elements and points are identical to those of the scheme in Subsection 3.1, as presented in Fig. \ref{fig_grid_stencil}. The stencil elements are
\begin{equation}
	S=\left\{\Omega_1,\Omega_2,\Omega_3,\Omega_4\right\}.
\end{equation}
The stencil points are
\begin{equation}
	Q_1^1=\left\{x_1,x_2,x_3\right\},\ \ Q_2^1=\left\{x_2,x_3,x_4\right\},\ \ Q_3^1=\left\{x_3,x_4,x_5\right\}.
\end{equation}
First, the symmetric stencil is considered, consisting of $S$ and $Q_2^1$. The scheme is expressed as follows, 
\begin{equation}
	\bm{f}\cdot\bm{c}=R\left(O\left(h^4\right)\right),
	\label{eq.3.24}
\end{equation}
where
\begin{equation} 
	\bm{c}=\left[c_1,c_2,c_3,\ \ldots,c_7\right]^T,\ \ \  \bm{f}=\left[{\bar{u}}_1,\ {\bar{u}}_2,\ {\bar{u}}_3,\ {\bar{u}}_4,h_2u_2^\prime,\ h_3u_3^\prime,\ h_4u_4^\prime\right].
\end{equation}
Eq. (\ref{eq.3.24}) holds exactly for the basis functions  $\bm{\varphi}\left(x\right)=\left[1,x,x^2,x^3\right]$, leading to the following system, i.e.,
\begin{equation}
	\mathbb{A}\bm{c}=0,
\end{equation}
where 
\begin{equation}
	\mathbb{A}=\left[\begin{matrix}1&1&1&1&0&0&0\\\frac{x_2^2-x_1^2}{{2h}_1}&\frac{x_3^2-x_2^2}{2h_2}&\frac{x_4^2-x_3^2}{{2h}_3}&\frac{x_5^2-x_4^2}{2h_4}&h_2&h_3&h_4\\\frac{x_2^3-x_1^3}{{3h}_1}&\frac{x_3^3-x_2^3}{3h_2}&\frac{x_4^3-x_3^3}{3h_3}&\frac{x_5^3-x_4^3}{{3h}_4}&2h_2x_2&2h_3x_2&2h_4x_2\\\frac{x_2^4-x_1^4}{{4h}_1}&\frac{x_3^4-x_2^4}{4h_2}&\frac{x_4^4-x_3^4}{4h_3}&\frac{x_5^4-x_4^4}{{4h}_4}&3h_2x_2^2&3h_3x_2^2&3h_4x_2^2\\\end{matrix}\right].
\end{equation}

Subsequently, the general solution for the coefficient vector $\bm{c}$ of this underdetermined homogeneous linear system is derived, following the same form as Eq. (\ref{eq_sub2.6}). For uniform grids, the general solution matrix $\mathbb{C}$ for the $\text{CFVM-}P(4)\text{-}S(4)\text{-}Q^1_2(3)$ scheme is given by
\begin{equation}
	\mathbb{C}=\frac{1}{12} \left(\begin{array}{ccccccc} 11 & -9 & -3 & 1 & 12 & 0 & 0\\ -1 & 15 & -15 & 1 & 0 & 12 & 0\\ -1 & 3 & 9 & -11 & 0 & 0 & 12 \end{array}\right)^T.
\end{equation}
The truncation error of this scheme at specific points can be expressed in the form of Eq. (\ref{eq_sub2.7}).

For points $x_i$ (where $i=2,3,4$), the corresponding truncation error matrices $Ri$ are
\begin{equation}
	\mathbb{R}_2=\left(\begin{array}{ccc} 2 & 0 & -2\\ \frac{19}{3} & \frac{4}{3} & -\frac{41}{3}\\ 20 & 8 & -64 \end{array}\right),\ \ 
	\mathbb{R}_3=\left(\begin{array}{ccc} 2 & 0 & -2\\ -\frac{11}{3} & \frac{4}{3} & -\frac{11}{3}\\ 12 & 0 & -12 \end{array}\right),\ \ 
	\mathbb{R}_4=\left(\begin{array}{ccc} 2 & 0 & -2\\ -\frac{41}{3} & \frac{4}{3} & \frac{19}{3}\\ 64 & -8 & -20 \end{array}\right).
\end{equation}
Since the point stencils $Q_1^1$ or  $Q_3^1$ are symmetric, $Q_1^1$ is used as an example herein. Similarly, for the scheme using the point stencil $Q^1_1$, the general solution matrix $\mathbb{C}$ is given by
\begin{equation}
	\mathbb{C}=\frac{1}{12}
	\left(\begin{array}{ccccccc} 35 & -69 & 45 & -11 & 12 & 0 & 0\\ 11 & -9 & -3 & 1 & 0 & 12 & 0\\ -1 & 15 & -15 & 1 & 0 & 0 & 12 \end{array}\right)^T.
\end{equation}
For points $x_i, i=1,2,3$, the corresponding truncation error matrices $Ri$ are
\begin{equation}
	\mathbb{R}_1=\left(\begin{array}{ccc} -20 & 2 & 0\\ -\frac{476}{3} & \frac{49}{3} & \frac{4}{3}\\ -860 & 88 & 16 \end{array}\right),\ \ 
	\mathbb{R}_2=\left(\begin{array}{ccc} -20 & 2 & 0\\ -\frac{176}{3} & \frac{19}{3} & \frac{4}{3}\\ -208 & 20 & 8 \end{array}\right),\ \ 
	\mathbb{R}_3=\left(\begin{array}{ccc} -20 & 2 & 0\\ \frac{124}{3} & -\frac{11}{3} & \frac{4}{3}\\ -156 & 12 & 0 \end{array}\right).
\end{equation}

\subsection{Hybrid schemes}
The preceding sections describe the construction of finite volume schemes based on element mean values, which utilise either function values or derivative values. Additionally, hybrid schemes can be constructed using values of different orders of derivatives at  points.  This enables the development of ultra-compact schemes, which offers the advantage of higher accuarcy within a minimal stencil.

The stencil for hybrid schemes comprises mean values of elements $S={\Omega_1, \Omega_2, \ldots, \Omega_s}$ and $m$-th order derivatives of the function at points $Q^m=\left\{q_1^m, q_2^m, \ldots, q_{n_m}^m \right\}$ (where $m=0,1,\cdots,M$). A $(k+1)$-th order compact finite volume scheme constructed using stencil $S$ and points $Q^m$ (for $m=0,1,2,\ldots,M$) is denoted as $CFVM\text{-}P(k+1)\text{-}S(s)\text{-}Q^1(n_1)\text{-}Q^2(n_2)\text{-}\ldots\text{-}Q^M(n_M)$. Here, a fifth-order compact scheme is constructed, denoted as $\text{CFVM-}P(5)\text{-}S(2)\text{-}Q^0(3)\text{-}Q^1(3)$.

 The stencil elements are
\begin{equation}
	S=\left\{\Omega_1,\Omega_2\right\}.
\end{equation}
The stencil points for $Q^0$ (function values) and $Q^0$ (first derivatives) are given as follow
\begin{equation}
	Q^0=Q^1=\left\{x_1,x_2,x_3\right\}.
\end{equation}
The varialbe vector $\bm{f}$  is defined as 
\begin{equation}
	\bm{f}=\left[{\bar{u}}_1,\ {\bar{u}}_2,u_1,u_2,u_3,h_1u_1^\prime,\ h_2u_2^\prime,\ h_3u_3^\prime\right].
\end{equation}
Here, $h_i=x_{i+1}-x_i$ denotes the length of element $\Omega_i$. For uniform grids, $h_1=h_2=h_3=h$.  Eq. (\ref{eq2.3}) holds exactly for the basis functions of a fourth-order polynomial, given by $\bm{\varphi}\left(x\right) = \left[1, x, x^2, x^3, x^4\right]^T$. From Eqs. (\ref{eq2.10}) and (\ref{eq2.10.new}), the expression for matrix $\mathbb{A}$ can be derived. Solving Eq. (\ref{eq2.9}) then yields the general solution for the coefficient vector $\bm{c}$. For uniform grids, the general solution matrix $\mathbb{C}$ (composed of basis solutions) is given by
\begin{equation}
	\mathbb{C}=\frac{1}{2}
	\left(\begin{array}{cccccccc} -23 & -7 & 12 & 16 & 2 & 2 & 0 & 0\\ 4 & -4 & -1 & 0 & 1 & 0 & 2 & 0\\ 7 & 23 & -2 & -16 & -12 & 0 & 0 & 2 \end{array}\right)^T.
\end{equation}
The truncation error of this scheme at points $x_i$ (for $i=1,2,3$) can be expressed as
\begin{equation}
	R_{x_i}=\left[\frac{u^{\left(5\right)}\left(x_i\right)}{5!}h^5,
	\frac{u^{\left(6\right)}\left(x_i\right)}{6!}h^6,
	\frac{u^{\left(7\right)}\left(x_i\right)}{7!}h^7
	\right]\mathbb{R}_i\bm{\eta}.
\end{equation}
For points $x_1$ to $x_3$ in uniform grids, the corresponding truncation error matrices $\mathbb{R}_i$ are
\begin{equation}
	\mathbb{R}_1=\left(\begin{array}{ccc} 
		\frac{4}{3} & \frac{1}{3} & \frac{4}{3}\\ 
		\frac{48}{7} & 2 & \frac{64}{7}\\ 
		23 & \frac{15}{2} & 39 \end{array}\right),\ \ 
	\mathbb{R}_2=\left(\begin{array}{ccc} 
		\frac{4}{3} & \frac{1}{3} & \frac{4}{3}\\ 
		-\frac{8}{7} & 0 & \frac{8}{7}\\ 
		3 & \frac{1}{2} & 3 \end{array}\right),\ \ 
	\mathbb{R}_3=\left(\begin{array}{ccc} 
		\frac{4}{3} & \frac{1}{3} & \frac{4}{3}\\ -\frac{64}{7} & -2 & -\frac{48}{7}\\ 39 & \frac{15}{2} & 23 \end{array}\right).
\end{equation}

\subsection{Compact schemes for triangular grids}
The construction of one-dimensional schemes was introduced earlier. Notably, the construction procedure for the aforementioned schemes can be relatively easily extended to the development of compact schemes for multidimensional unstructured grids. To illustrate this, a fourth-order accurate compact scheme for triangular unstructured grids is constructed in this section using the method proposed herein.

\begin{figure}[!t]
	\centering
	\includegraphics[height=5.5cm]{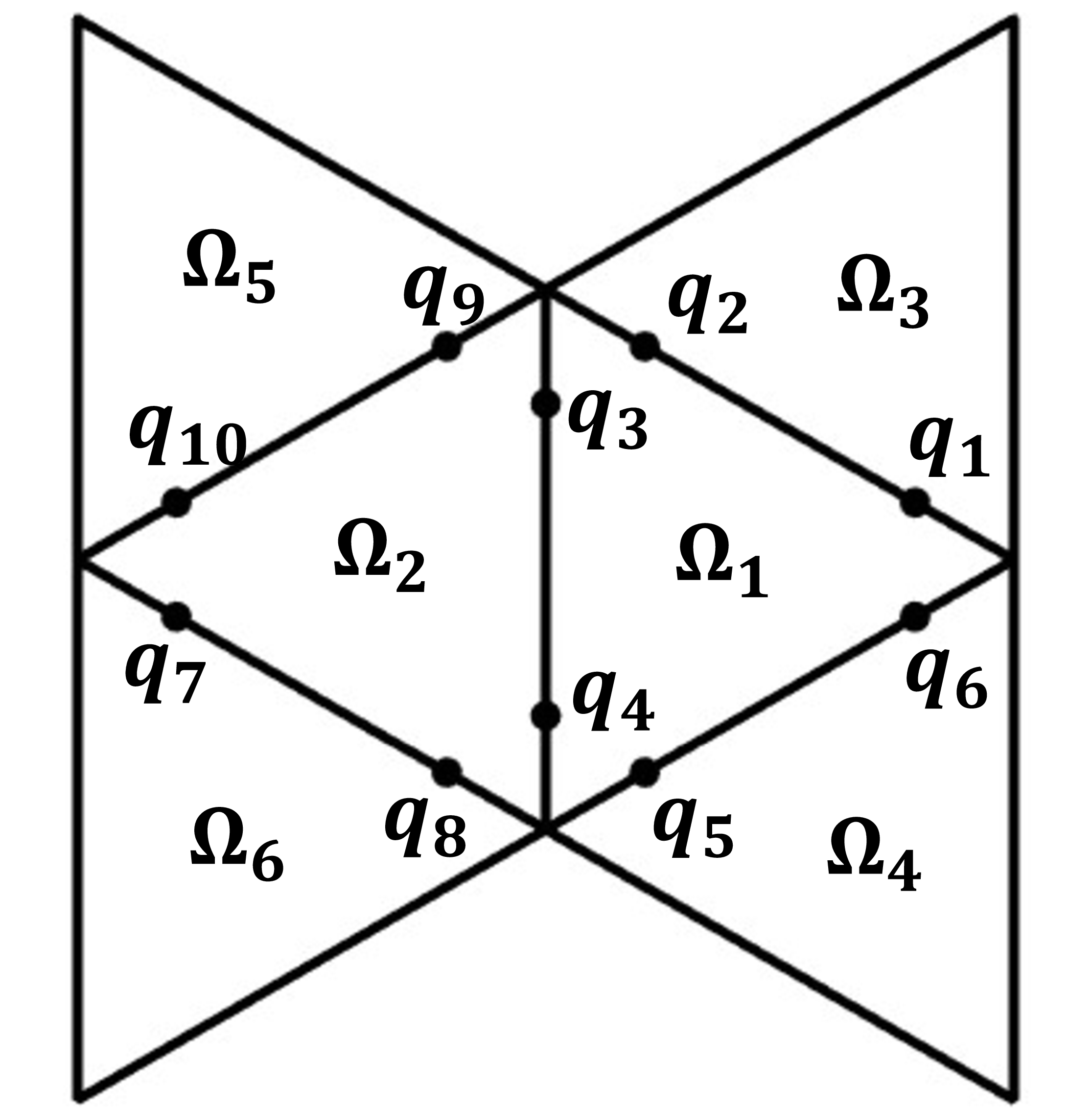}
	\caption{Stencil points and elements of the fourth-order CFVM scheme for unstructured triangular grids.}
	\label{fig_stencil_2D}
\end{figure}

Fig. \ref{fig_stencil_2D} presents the stencil used for the fourth-order accurate compact scheme on unstructured triangular grids. 
For a given edge, the elements on both sides of the edge and their adjacent elements are selected as stencil elements, denoted as $S$, i.e.,
\begin{equation}
	S=\left\{\Omega_1,\Omega_2,\Omega_3,\Omega_4,\Omega_5,\Omega_6\right\}.
	\label{eq_sub4.1}
\end{equation}
Two Gaussian integration points are introduced on each edge, and a total of 10 points on the edges of adjacent elements are selected as stencil points $Q$, i.e.,
\begin{equation}
	Q=\left\{q_1,q_2,q_3,\ldots,q_{10}\right\}.
	\label{eq_sub4.2}
\end{equation}
The constructed scheme follows the form of Eq. (\ref{eq2.3}), i.e.,
\begin{equation}
	\bm{f}\cdot\bm{c}=R\left(O\left(h^4\right)\right),
	\label{eq_sub4.3}
\end{equation}
where
\begin{equation} 
	\bm{c}=\left[c_1,c_2,c_3,\ldots,c_{16}\right]^T,\ \  \bm{f}=\left[{\bar{u}}_1,{\bar{u}}_2,\ldots {\bar{u}}_6,u_{q_1},u_{q_2},\ldots u_{q_{10}}\right].
\end{equation}
The scheme achieving fourth-order accuracy is equivalent to Eq. (\ref{eq_sub4.3}) holding exactly for any two-dimensional cubic polynomial.  This means  $\bm{f}\cdot\bm{c}=0$   when the function distribution corresponds to any of the basis functions of a two-dimensional cubic polynomial, denoted as $\bm{\varphi}$. The expression of $\bm{\varphi}$ is
\begin{equation}
	\bm{\varphi}\left(x,y\right)=\left[1,x,y,x^2,xy,y^2,x^3,x^2y,xy^2,y^3\right]^T.
	\label{eq_sub4.5}
\end{equation}
Substituting the basis functions $\bm{\varphi}$ into Eq. (\ref{eq_sub4.3}) yields an underdetermined homogeneous linear system, i.e.,
\begin{equation}
	\mathbb{A}\bm{c}=0,
	\label{eq_sub4.6}
\end{equation}
where $\mathbb{A}$ is a $10\times 16$ matrix. The $i$-th column of the matrix $\mathbb{A}$ is defined by $\bm{a}_i$ (for $i=1,2,\ldots,16$). The expression for $\bm{a}_i$ is given by
\begin{equation}
	\begin{aligned}
		& \bm{a}_i=\frac{1}{s_i}\int \int_{\Omega_i}\bm{\varphi}\left(x\right)dxdy,\ \ i=1,2,\ldots,6;\\
		& \bm{a}_{i+6}=\bm{\varphi}\left(q_{i}\right),\ \ i=1,2,\ldots,10.
		\label{eq_sub4.7}
	\end{aligned}
\end{equation}
Subsequently, the general solution for $\bm{c}$  is derived as
\begin{equation}
	\bm{c}=\eta_1\bm{c}_1+\eta_2\bm{c}_2+\ldots+\eta_{k}\bm{c}_{k}=\mathbb{C}\bm{\eta},
	\label{eq_sub4.8}
\end{equation}
where $\bm{c}_i$ (columns of $\mathbb{C}$) are the basis solutions of Eq. (\ref{eq_sub4.7}), $k=6$ is the number of basis solutions (dimension of the null space), $\eta_i$ (for $i=1,2,\ldots ,k$) are arbitrary coefficients, and $\mathbb{C}$ is an $16\times 6$ matrix composed of the basis solutions $\bm{c}_i$. Any linear combination of these basis vectors within this space gets fourth-order accuracy.

Thus far, a fourth-order accurate scheme space for two-dimensional unstructured triangular grids has been successfully constructed. In reconstruction,  we only need to select an appropriate combination in this space, that is, select coefficients $\bm{\eta}$, to obtain a scheme with fourth-order accuracy for the unstructured grids. Similarly, compact schemes for approximating derivatives on unstructured triangular grids can be constructed using the same procedure. In Subsection 4.5, an  example of the two-dimensional scalar linear advection equation is presented to verify the accuracy of the proposed scheme.

\section{Numerical results}
In this section, several  benchmark tests are conducted to evaluate the accuracy, robustness, and shock-capturing capability of the proposed compact finite volume schemes.

\subsection{The scalar linear advection equation}
In this subsection, a set of tests on the scalar linear convection equation is employed to verify the properties of the aforementioned schemes. The governing equation is given by
 \begin{equation}
 	\frac{\partial u}{\partial t}+a\frac{\partial u}{\partial x}=0,
 \end{equation}
where  $a$ is a real constant. In the tests, $a=1$ is adopted.  The computational domain is defined as $0 \le x\le 1$. The fourth-order Runge-Kutta method is used for time discretization, and periodic boundary conditions are applied. For the initial condition $u(x,t=0)=u_0(x)$, the exact solution is $u(x,t) = u_0(x-at)$.

The tests  consist of  two parts. The first part uses a sine function as the test function to evaluate the scheme’s accuracy for unsteady problems. The initial condition here is $u_0(x)=\sin\left(2 \pi kx\right)$, where $k$ is a positive integer  representing  the wave number of the sine wave.  The exact solution is  $u(x,t)=\sin\left(2\pi k\left(x-at\right)\right)$.

\begin{figure}[!t]
	\centering
	\includegraphics[height=6.5cm]{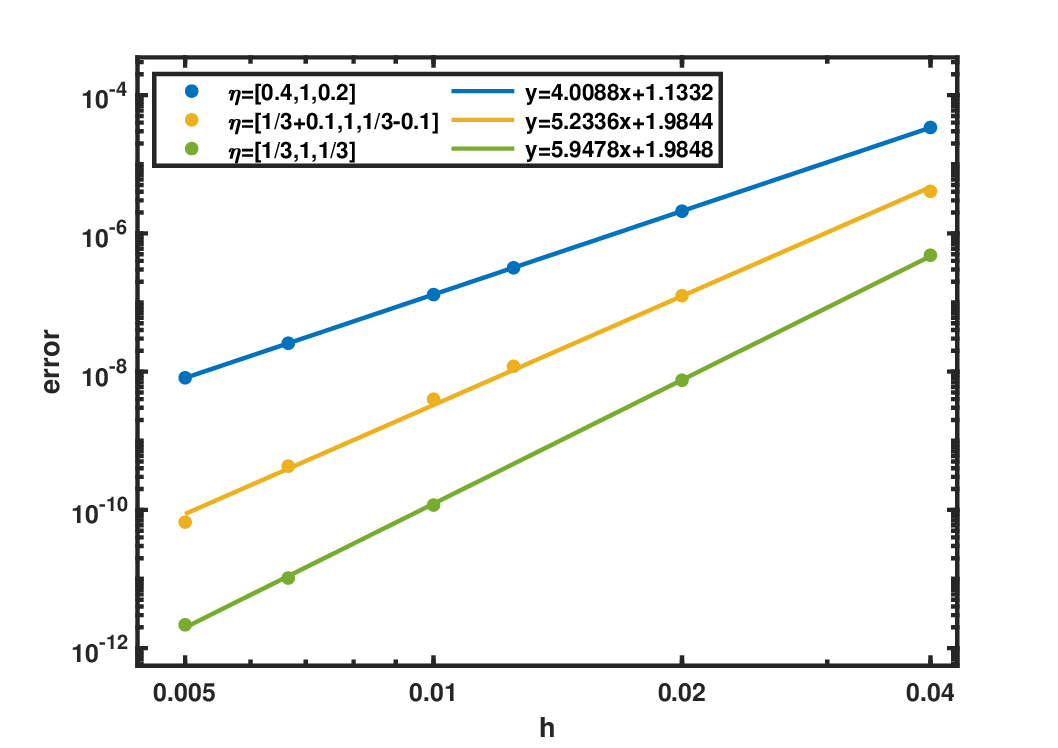}
	\caption{ $L_1$ error for different grid sizes in solving the linear advection equation.}
	\label{fig_linear_accuarcy}
\end{figure}

The $\text{CFVM-}P(4)\text{-}S(4)\text{-}Q^0_2(3)$ scheme is employed in this test, and the computed results are compared with the exact solution at $t=1$. Fig. \ref{fig_linear_accuarcy} presents the $L_1$ errors for different grid sizes. The slope of the fitted line represents the order of accuracy of the scheme. As shown in Fig. \ref{fig_linear_accuarcy}, for the parameters $\bm{\eta} = [0.4,1,0.2]^T$, $\bm{\eta} = [1/3+0.1,1,1/3-0.1]^T$ and  $\bm{\eta}=[1/3,1,1/3]^T$,  the numerical orders of accuracy (obtained from the fitted lines) are 4.0, 5.2, and 5.9, respectively.
These results are consistent with the theoretical orders of accuracy (4th, 5th, and 6th), 
 
\begin{figure}[h]
	\centering
	\begin{subfigure}[b]{0.45\textwidth} 
		\centering
		\includegraphics[height=5cm]{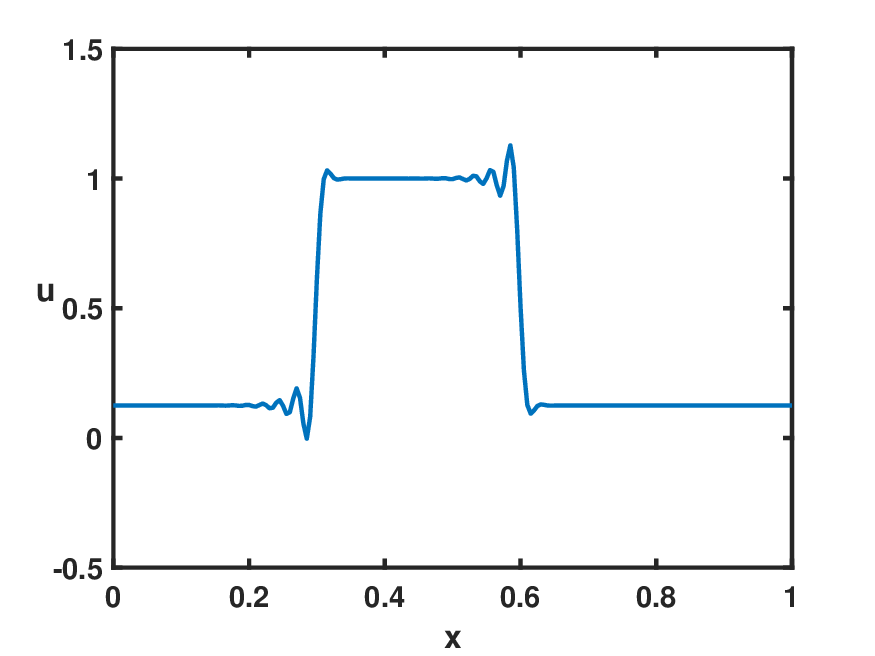}
		\caption{$\bm{\eta}=[0.4,1,0.2]^T$} 
		\label{fig_eta1}
	\end{subfigure}
	\hspace{0.2in} 
	\begin{subfigure}[b]{0.45\textwidth}
		\centering
		\includegraphics[height=5cm]{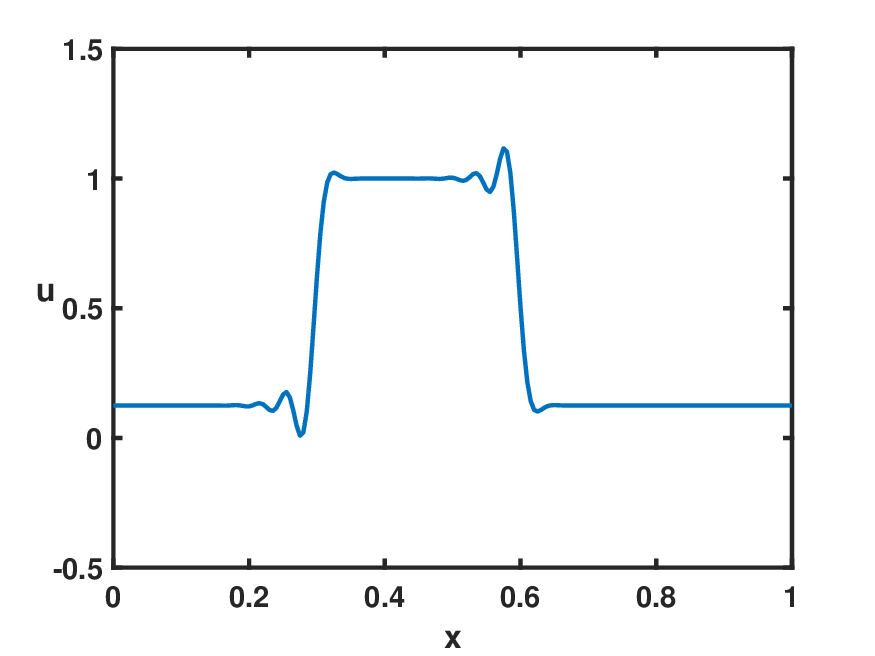}
		\caption{$\bm{\eta}=[0.4,1,-0.4]^T$}
		\label{fig_eta2}
	\end{subfigure}
	\hspace{0.2in}
	\begin{subfigure}[b]{0.45\textwidth}
		\centering
		\includegraphics[height=5cm]{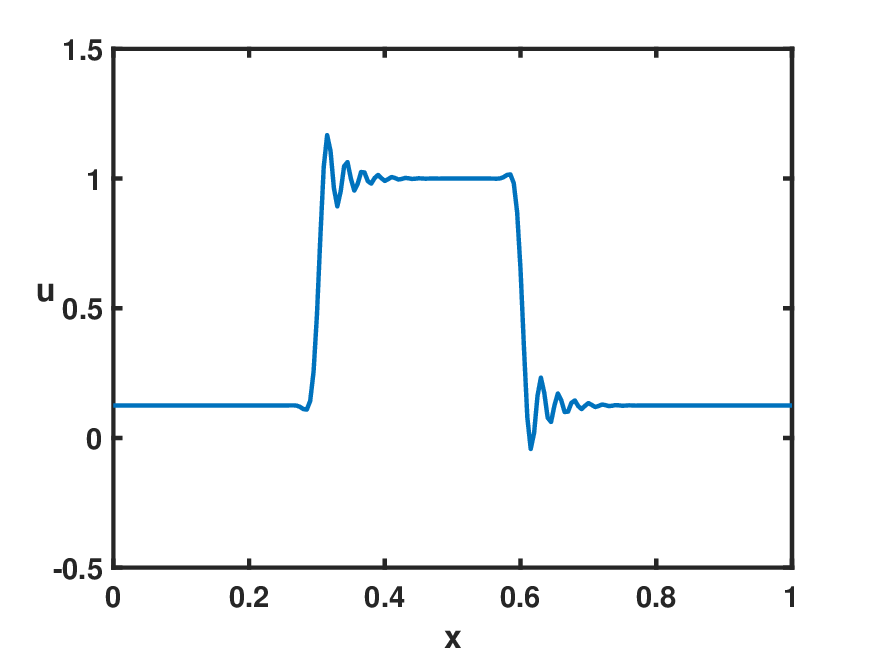}
		\caption{$\bm{\eta}=[0.5,1,0.3]^T$}
		\label{fig_eta3}
	\end{subfigure}
	\hspace{0.2in}
	\begin{subfigure}[b]{0.45\textwidth}
		\centering
		\includegraphics[height=5cm]{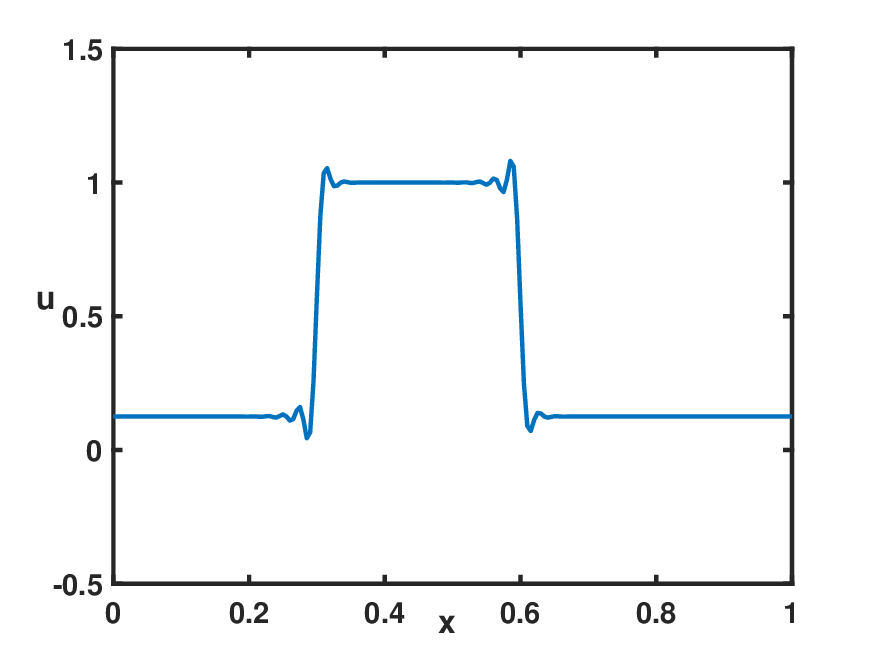}
		\caption{$\bm{\eta}=[1/3-0.1,1,1/3+0.1]^T$}
		\label{fig_eta4}
	\end{subfigure}
	\caption{Results for different $\bm{\eta}$ values in solving the linear advection equation with a square wave (one period).}
	\label{fig_different_eta_square}
\end{figure}

The second part uses a square wave as the test function to evaluate the dissipation and dispersion of the scheme and the WFVM scheme’s shock-capturing capability. The same $\text{CFVM-}P(4)\text{-}S(4)\text{-}Q^0_2(3)$ scheme is used in this test.

Fig. \ref{fig_different_eta_square} presents the results for different values of $\bm{\eta}$ in solving the linear advection equation with a square wave after one propagation period. The values of $\bm{\eta}$  in the four cases are $\bm{\eta}_a=[0.4,1,0.2]^T$, $\bm{\eta}_b=[0.4,1,-0.4]^T$, $\bm{\eta}_c=[0.5,1,0.3]^T$ and $\bm{\eta}_d=[1/3-0.1,1,1/3+0.1]^T$, respectively. From Eq. (\ref{eq_sub2.11.3}), it follows that for $\bm{\eta}_a$, the scheme exhibits $\frac{\omega_r}{ak}<1$, i.e., slow dispersion. In Fig. \ref{fig_eta1}, numerical oscillations are mainly observed on the left side of the discontinuity. This confirms that for $\bm{\eta}_a$, the wave propagation speed is lower than the exact value, which is consistent with the theoretical prediction. Similarly, for $\bm{\eta}_c$, the scheme exhibits $\frac{\omega_r}{ak}>1$, i.e., fast dispersion. As shown in Fig. \ref{fig_eta3}, numerical oscillations are mainly located on the right side of the discontinuity. For $\bm{\eta}_b$, the scheme exhibits zero numerical dissipation. Consequently, the dissipation effect on waves of all frequency bands is negligible, leading to numerical oscillations spreading across the entire computational domain. For $\bm{\eta}_c$, the scheme exhibits larger numerical dissipation compared to the other cases.  This conclusion is confirmed by Fig. \ref{fig_eta4} , where more high-frequency oscillations are suppressed and oscillations are smaller than those in the other three cases. These results demonstrate that the dispersion and dissipation characteristics of the compact scheme can be controlled by selecting different values of $\bm{\eta}$.

\begin{figure}[!t]
	\centering
	\includegraphics[height=5cm]{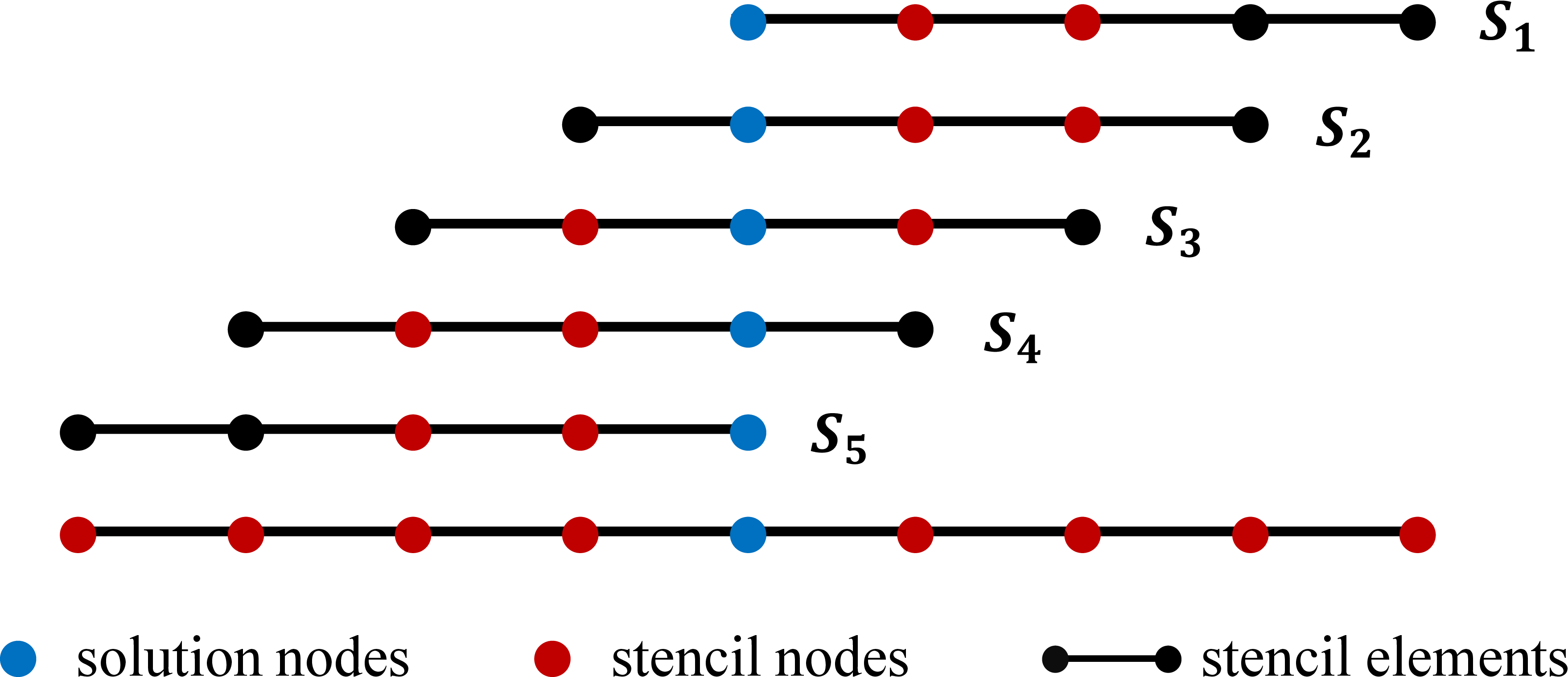}
	\caption{Five Sub-Stencils Used in the WCFV Scheme.}
	\label{fig_WFVM_stencil}
\end{figure}

Building on the above tests, the square wave advection problem is solved using the proposed weighted compact finite volume (WCFV) scheme. The WCFV scheme used herein consists of five sub-stencils, as presented in Fig. \ref{fig_WFVM_stencil}. Each sub-stencil employs the $\text{CFVM-}P(4)\text{-}S(4)\text{-}Q^0(3)$ scheme. Details of this scheme are provided in Subsection 3.1. For sub-stencils $S_1$ to $S_5$, the corresponding point stencils are $\left(Q_1^0,Q_2^0,Q_2^0,Q_2^0,Q_3^0\right)$, respectively. The corresponding coefficient vectors are $\bm{\eta}_{S_1}=[1,0,-1]^T$, $\bm{\eta}_{S_2}=[1,0,1]^T$, $\bm{\eta}_{S_3}=[1/3+0.001,0,1/3-0.001]^T$ ,$\bm{\eta}_{S_4}=[1,0,1]^T$ and $\bm{\eta}_{S_5}=[-1,0,1]^T$,  respectively. The contribution of each sub-stencil in the WCFV scheme is indicated by the weight $d_r$, as defined in Eq. (\ref{Eq4.9}). In this test, the values of dr for these five sub-stencils are set to $(0.0002,0.2,10,1,1)$.

\begin{figure}[!t]
	\centering
	\begin{subfigure}[b]{0.45\textwidth} 
		\centering
		\includegraphics[height=5cm]{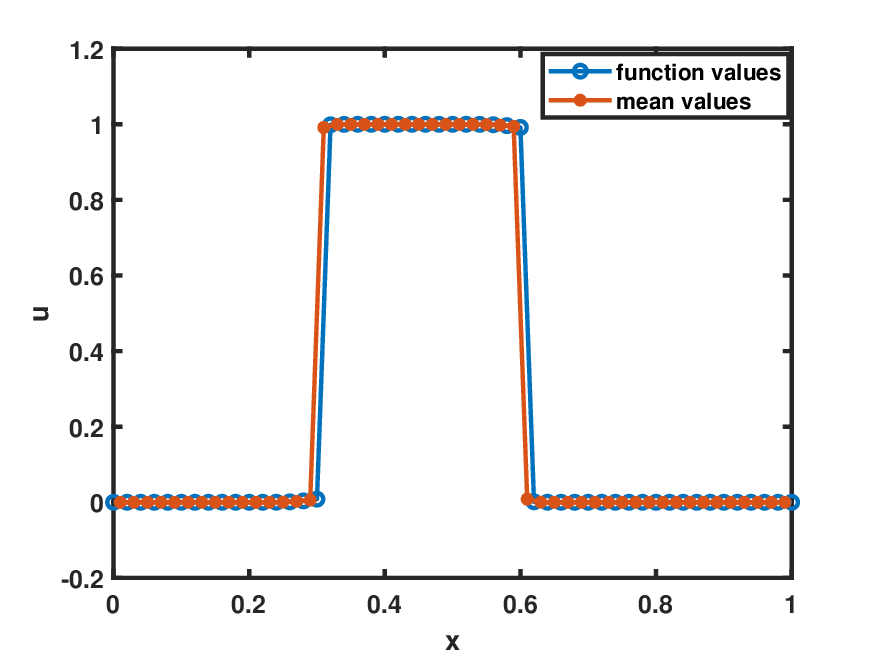}
		\caption{N=50} 
		\label{fig_square.a} 
	\end{subfigure}
	\hspace{0.2in} 
	\begin{subfigure}[b]{0.45\textwidth}
		\centering
		\includegraphics[height=5cm]{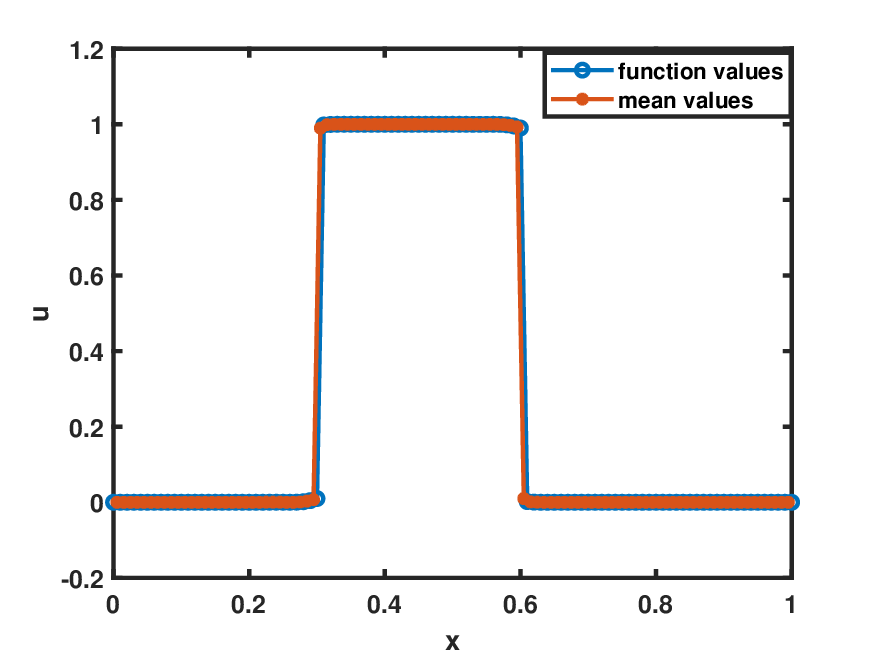}
		\caption{N=100} 
		\label{fig_square.b}
	\end{subfigure}
	\caption{WCFV scheme results for square wave advection at one period.}
	\label{fig_square} 
\end{figure}

Fig. \ref{fig_square} presents the simulation results for $N=50$ and $N=100$ (grid numbers) using the WCFV scheme. It can be observed that oscillations at the discontinuity are well suppressed, and sharp discontinuities are accurately captured.

\subsection{Burgers equation}
The Burgers equation is a nonlinear partial differential equation (PDE) with a known exact solution. The purpose of this test is to verify the accuracy of the proposed CFVM schemes when solving equations with nonlinear terms.

The Burgers equation is given by
\begin{equation}
	\frac{\partial u}{\partial t}+u\frac{\partial u}{\partial x}=\nu\frac{\partial^2u}{\partial x^2},
	\label{eq_bur1}
\end{equation}
where $\nu$ is a positive constant representing the viscosity coefficient. Integrating Eq. (\ref{eq_bur1}) over element $\Omega_i$ yields the semi-discrete finite volume formulation, i.e.,
\begin{equation}
	\frac{\partial{\bar{u}}_i}{\partial t}+\frac{1}{h}\left(u_{i+1}^2-u_i^2\right)=\frac{\nu}{h}\left[\left(\frac{\partial u}{\partial x}\right)_{i+1}-\left(\frac{\partial u}{\partial x}\right)_{i}\right].
	\label{eq_bur2}
\end{equation}
The treatment of the convective flux $u_{i}^2$ in Eq. (\ref{eq_bur2}) is detailed in Subsection 2.3. The $\text{CFVM-}P(4)\text{-}S(4)\text{-}Q^0_2(3)$ scheme with $\bm{\eta}=(0.3,1,0.3)$ is employed for the convective flux calculation. For the viscous flux term $\left(\frac{\partial u}{\partial x}\right)_{i}$, the first-order derivative is computed using the  mean values $\bar{u}_i$. The specific scheme for derivative approximation is detailed in Subsections 3.2 (the first derivative approximation schemes) and 3.3 (hybrid schemes). In this test, the $\text{CFVM-}P(4)\text{-}S(4) \text{-}Q^1_2(3)$  scheme (first derivative approximation) with $\bm{\eta}=(0.4,1,0.4)$ is used for the viscous flux calculation.

For the initial condition $ u_0\left(x\right) = g\left(x\right)$, the exact solution is given by the Cole-Hopf transformation, i.e.,
\begin{equation}
	u\left(x,t\right)=\frac{\int_{-\infty}^{\infty}{\frac{x-\eta}{t}e^{-\frac{G}{2\nu}}}d\eta}{\int_{-\infty}^{\infty}e^{-\frac{G}{2\nu}}d\eta},
	\label{eq_bur4}
\end{equation}
where the function $G$ is defined as
\begin{equation}
	G\left(\eta,x,t\right)=\int_{0}^{\eta}g\left(\widetilde{\eta}\right)d\widetilde{\eta}+\frac{\left(x-\eta\right)^2}{2t}.
	\label{eq_bur5}
\end{equation}
Specifically, the initial condition in this test is $u_0\left(x\right)=\sin\left(x\right)$, and the computational domain is $x\in\left[0,2\pi \right]$ with periodic boundary conditions. The fourth-order Runge-Kutta method is used for time discretization, as in Subsection 4.1. For the initial condition $u_0\left(x\right)=\sin\left(x\right)$, the function $G\left(\eta, t\right)$ in the exact solution simplifies to $1 - \cos\left(\eta\right) + \left(x - \eta\right)^2/2t$. Substituting this expression for $G\left(\eta, t\right)$ into Eq. (\ref{eq_bur5}) yields the exact solution for the Burgers equation in this test.

\begin{figure}[!t]
	\centering
	\begin{subfigure}[b]{0.45\textwidth} 
		\centering
		\includegraphics[height=5cm]{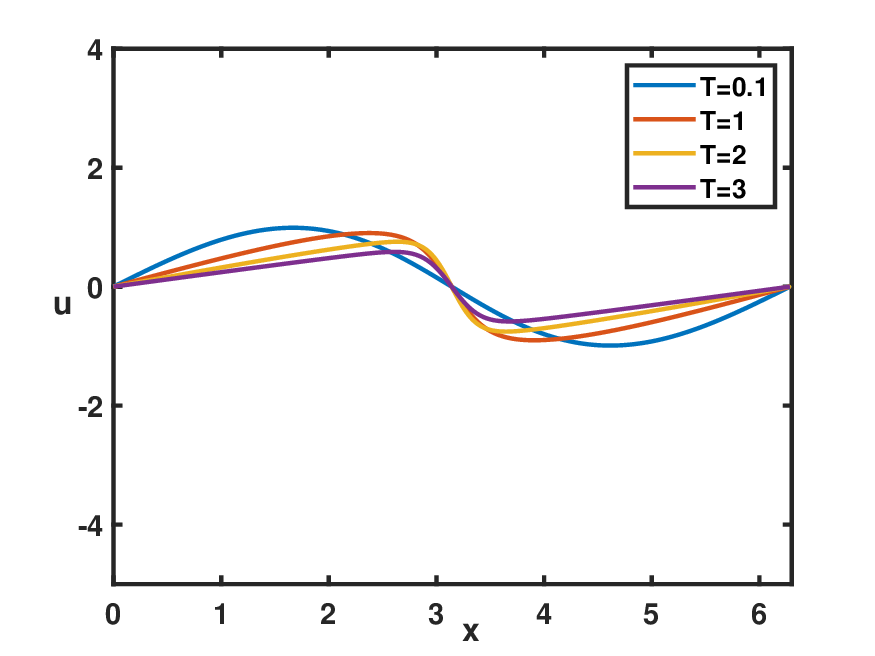}
		\caption{The results at different time} 
		\label{fig_burgers_solution} 
	\end{subfigure}
	\hspace{0.2in} 
	\begin{subfigure}[b]{0.45\textwidth}
		\centering
		\includegraphics[height=5cm]{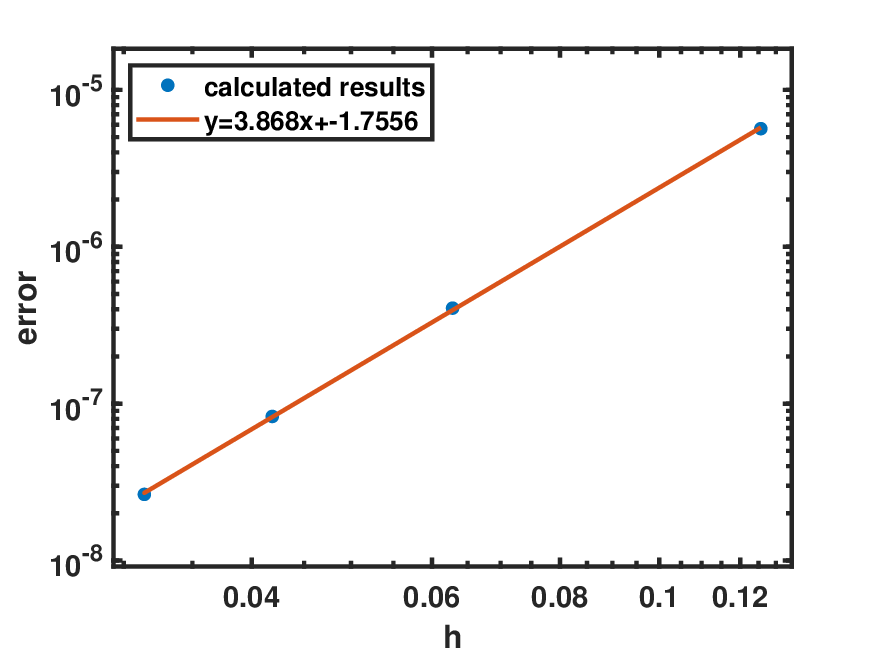}
		\caption{$L_1$ error within different grid sizes.} 
		\label{fig_burgers_accuarcy}
	\end{subfigure}
	\caption{Numerical solutions of the Burgers equation at different times (left) and  $L_1$ error within different grid size at $t=1.0$ (right).}
	\label{fig_burgers} 
\end{figure}

Fig. \ref{fig_burgers_solution} presents the numerical solutions of the Burgers equation at different time instants ($T=0.1,1,2,3)$, and Fig. \ref{fig_burgers_accuarcy} shows the $L_1$ errors for different grid sizes at $t=1.0$. From Fig. \ref{fig_burgers_accuarcy}, the numerical order of accuracy (slope of the $L_1$-error curve) is 3.87, which is consistent with the theoretical fourth-order accuracy.

\subsection{Shock tube problem}

In this subsection, shock tube problems, specifically the Sod problem, are solved to evaluate the ability of the WCFV scheme to capture strong shock waves and  discontinuities.

The initial conditions for the Sod problem are given by
\begin{equation}
	\left(u_0, \rho_0,  p_0\right)= 
	\begin{cases} 
		(0.698,0.445,3.528) & \text{for } -1\leq x \leq 0,\\ 
		(0,0.5,0.571)      & \text{for } 0< x \leq 1.
\end{cases}
\end{equation}
The computational domain is $\left[-1,1\right]$, which is discretized into $N=400$ uniform grid elements. The time discretisation scheme is the fourth-order Runge-Kutta method and the Courant number is CFL=1.0. The numerical results for the Sod problem are computed at  $T=0.2$.

In this test, the WCFV scheme is employed to handle discontinuities, as described in Subsection 2.3.
The stencil configuration is identical to that in Test 4.1 (see Fig. 13). The coefficient vectors $\bm{\eta}$ and the stencil weights $d_r$ for calculating left-going fluxes are
\begin{equation}
\begin{aligned}
 &\bm{\eta}_{S_1}=[1,0,-1],\ \bm{\eta}_{S_2}=[1,0,1],\  \bm{\eta}_{S_3}=[\frac{1}{3},0,\frac{1}{3}],\ 
 \bm{\eta}_{S_4}=[1,0,1],\  \bm{\eta}_{S_5}=[-1,0,1],\\
 &d_r=(0.0002,0.2,10,1,1).
\end{aligned}
\end{equation}
For right-going fluxes, the coefficient vectors $\bm{\eta}$  and stencil weights $d_r$ are
\begin{equation}
	\begin{aligned}
		&\bm{\eta}_{S_1}=[-1,0,1],\ \bm{\eta}_{S_2}=[1,0,1],\  \bm{\eta}_{S_3}=[\frac{1}{3},0,\frac{1}{3}],\ 
		\bm{\eta}_{S_4}=[1,0,1],\  \bm{\eta}_{S_5}=[1,0,-1],\\
		&d_r=(1,1,10,0.2,0.0002).
	\end{aligned}
\end{equation}

\begin{figure}[!t]
	\centering
	\includegraphics[height=6.5cm]{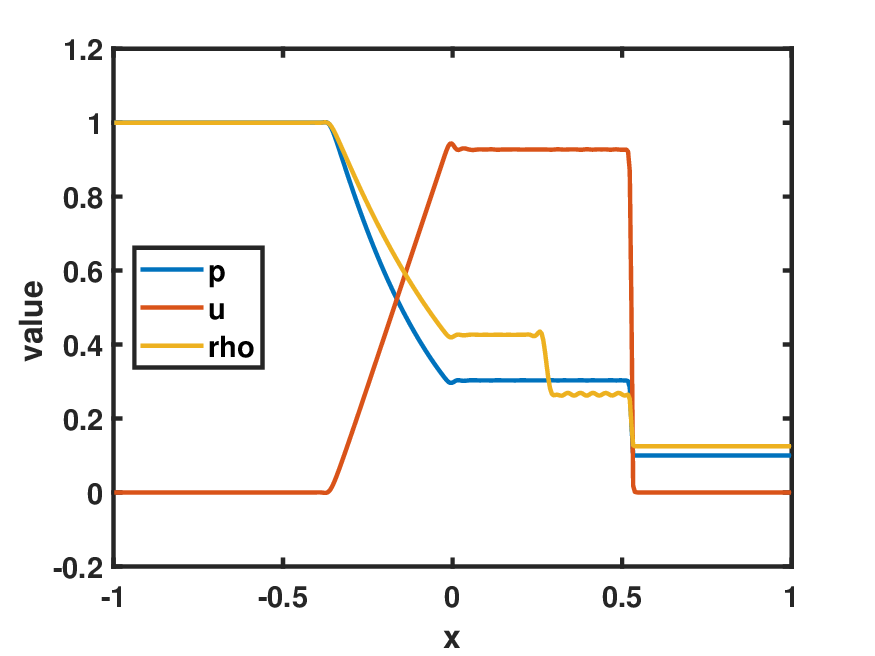}
	\caption{The numerical solution of the Sod problem at $T=0.2$.}
	\label{fig_sod_problem}
\end{figure}

Fig. \ref{fig_sod_problem} presents the distributions of density, velocity, and pressure, demonstrating that the WCFV scheme captures strong shock waves and contact discontinuities with minimal numerical oscillations.

\subsection{Shu-Osher problem}
The Shu-Osher problem features both strong discontinuities and multi-scale flow structures, making it a challenging test case for shock-capturing schemes.  This test case is used to evaluate the resolution and shock-capturing capability of the proposed WCFV method.

The initial condition involves a shock wave interacting with a sine wave, given by
\begin{equation}
	\left(u_0, \rho_0,  p_0\right)= 
	\begin{cases} 
		(2.629369,3.857143,10.3333)         & \text{for } 0\leq x \leq 1,\\ 
		(0,1+0.2\sin(5x),1)    & \text{for } 1< x \leq 10.
	\end{cases}
\end{equation}

The computational domain is $\left[0,10\right]$, discretized into $N=400$ uniform grid elements. The simulation is run until  $T=0.6$, using the fourth-order Runge-Kutta (RK4) method for time discretization. The WCFV scheme used herein is identical to that employed for the Sod problem (see Subsection 4.3).

\begin{figure}[!t]
	\centering
	\includegraphics[height=6.5cm]{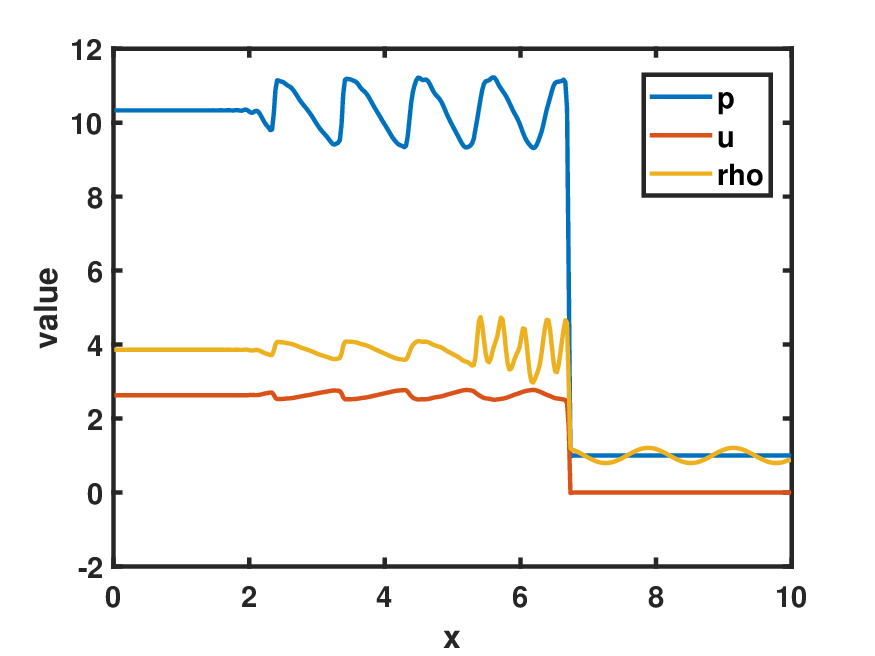}
	\caption{WCFV Scheme Results for the Shu-Osher Problem at $T=0.6$, $N=400$.}
	\label{fig_shu_osher_problem}
\end{figure}

Fig. \ref{fig_shu_osher_problem} presents the numerical results of the Shu-Osher problem obtained using the WCFV scheme. As shown in Fig. \ref{fig_shu_osher_problem}, the WCFV scheme accurately captures both multi-scale flow structures and strong discontinuities even on a relatively coarse grid (N=400).

\subsection{ The two-dimensional scalar linear advection equation}
The two-dimensional (2D) scalar linear advection equation is solved to verify the accuracy of the proposed CFVM scheme for unsteady 2D problems.

\begin{figure}[!t]
	\centering
	\begin{subfigure}[b]{0.45\textwidth} 
		\centering
		\includegraphics[height=5cm]{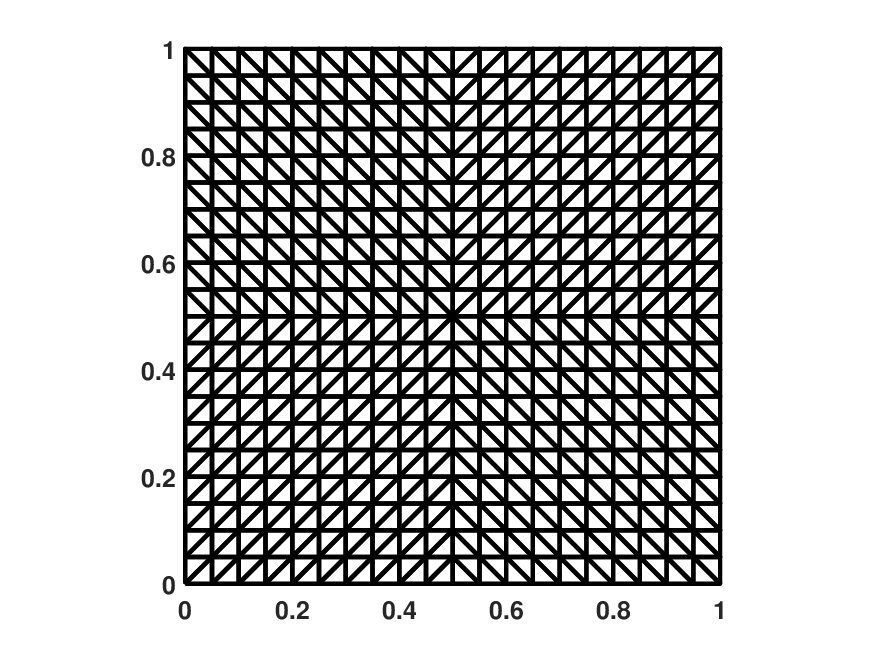}
		\caption{The unstructured triangular  grids.} 
		\label{fig_triangle_grid} 
	\end{subfigure}
	\hspace{0.05in} 
	\begin{subfigure}[b]{0.45\textwidth}
		\centering
		\includegraphics[height=5cm]{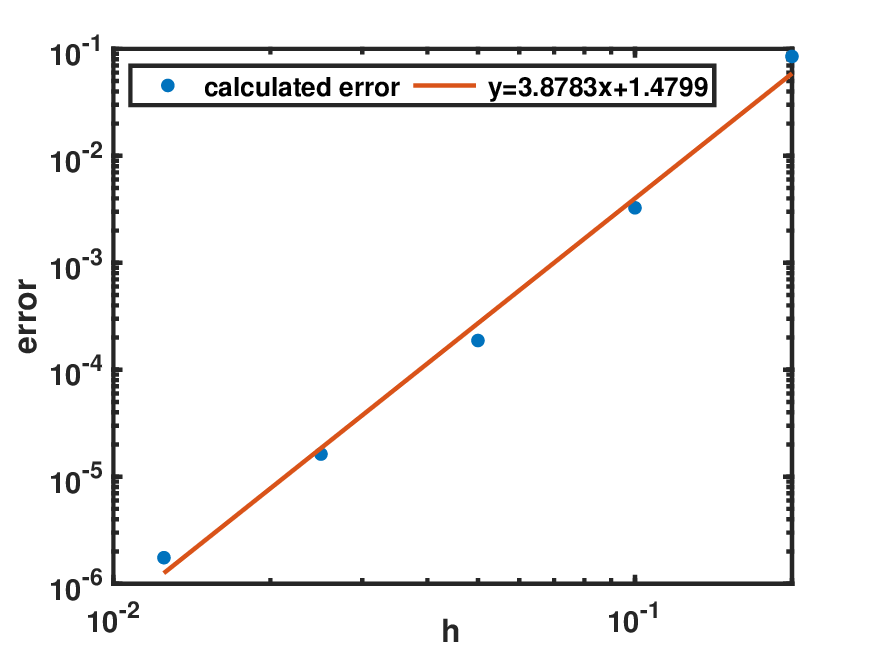}
		\caption{$L1$ error versus grid size.} 
		\label{fig_conv_2D_accuarcy}
	\end{subfigure}
	\caption{Unstructured triangular  grids and $L1$ error versus grid size for the 2D linear advection equation.}
	\label{fig_Kovasnay_flow} 
\end{figure}

The governing equation is
\begin{equation}
	\frac{\partial u}{\partial t}+a\frac{\partial u}{\partial x}+b\frac{\partial u}{\partial y}=0,
\end{equation}
where $(a,b)$ are the convective velocities in the $x$ and $y$ directions, respectively. In this test, the computational domain is $\Omega=\left\{\left(x,y\right)\middle|0\le x\le1,\ \ 0\le y\le1\right\}$, with periodic boundary conditions applied in both $x$ and $y$ directions. The right triangular unstructured grids used in this test are presented in Fig. \ref{fig_triangle_grid}. The exact solution of the 2D scalar linear advection equation is $u(x, y, t) = u_0(x - at, y - bt)$, where $u_0(x,y)$ is the initial condition. The convective velocities are set to $(a,b)=(-1,-1)$, and the initial condition is $u_0(x, y) = \sin(2\pi x)\cos(2\pi y)$. The classical fourth-order Runge-Kutta scheme (Subsection 2.3) is used for time discretization, and the 2D compact scheme from Subsection 3.4 is employed for spatial discretization. The simulation is run until $T=1.0$, after which the solution has completed one full period of advection.

Fig. 18b presents the $L_1$ errors for different grid sizes. The slope of the fitted line is 3.88, confirming that the proposed 2D CFVM scheme achieves the expected theoretical fourth-order accuracy. This test demonstrates that the proposed CFVM method can be extended to unstructured grids while maintaining the designed fourth-order accuracy.

\section{Conclusion}
This paper proposes a novel method for constructing compact finite volume schemes, with a focus on the detailed analysis of one-dimensional formulations. Compact schemes are constructed by establishing a linear approximation relationship between the mean values of stencil elements, and the function values and derivatives at stencil points. An approach equivalent to Taylor series expansion is employed to determine the coefficients of this linear approximation relationship, eliminating the need for direct Taylor expansion, which is cumbersome for unstructured grids. Specifically, a scheme meets the desired accuracy level if it holds exactly for the basis functions of polynomials of the corresponding degree (e.g., fourth-order accuracy requires exactness for cubic polynomials).
Through this procedure, the problem of constructing compact schemes is transformed into solving an underdetermined homogeneous linear system—a key insight that simplifies scheme construction.

Solving the null space of this system yields all schemes that meet the accuracy requirements and this set of schemes is termed the ‘scheme space’.  Schemes within the scheme space possess the same (or higher) accuracy but exhibit distinct dispersion and dissipation characteristics, enabling flexible control over these properties. Fourier analysis is used to characterize the dispersion and dissipation of each scheme in the space, providing a theoretical basis for parameter selection. Furthermore, the WCFV (Weighted Compact Finite Volume) scheme is developed by integrating the WENO concept with multiple stencils, enabling nonlinear oscillation suppression at discontinuities. One-dimensional benchmark tests (linear advection, Burgers equation, Sod shock tube, Shu-Osher problem) confirm that the proposed schemes achieve the designed accuracy and exhibit effective shock/discontinuity-capturing capabilities.

Notably, the scheme space construction method can be readily extended to multidimensional unstructured grids, addressing a key limitation of classical compact methods. This is because, for grids of any dimension, the scheme construction problem is transformed into solving a unified underdetermined homogeneous linear system via the proposed procedure—ensuring methodological consistency across 1D, 2D, and 3D grids. This implies that scheme construction for different dimensions involves identical mathematical operations, simplifying implementation for complex grids. To demonstrate this, a fourth-order accurate scheme space for 2D unstructured triangular grids is constructed in Section 3.4, and its accuracy is verified via an  two-dimensional scalar linear advection test in Section 4.5.

The application of this method on unstructured grids extends far beyond the examples presented herein. Several promising directions for future research are identified as follows.
First, construction of high-accuracy compact schemes for diverse unstructured grid types, including mixed-element grids, high-aspect-ratio grids, and grids with curved boundaries.  Besides, extension of the nonlinear weighted compact scheme (WCFV) to unstructured grids, along with the development of adaptive optimal dispersion/dissipation control strategies and efficient parallel computing implementations. Future work will focus on addressing these challenges to further advance the proposed method.

\section*{Acknowledgments}

Additional information can be given in the template, such as to not include funder information in the acknowledgments section.

\bibliography{refs}

\end{document}